\PassOptionsToPackage{dvipsnames}{xcolor}
\documentclass[twocolumn,trackchanges]{aastex631}
\usepackage{amstext,amsmath}
\usepackage[T1]{fontenc}
\usepackage[figure,figure*]{hypcap}
\usepackage{comment}
\usepackage{physics}

\usepackage[bb=boondox]{mathalfa}
\usepackage{graphicx}
\usepackage{multirow}
\usepackage{subfig}
\usepackage{colortbl}
\usepackage{float}

\shorttitle{Fitting Foreground Models}
\shortauthors{Hibbard et al.}

\graphicspath{{./}{figures/}}

\begin{document}

\title{Fitting and Comparing Galactic Foreground Models for Unbiased 21-cm Cosmology}

\correspondingauthor{Joshua J. Hibbard}
\author[0000-0002-9377-5133]{Joshua J. Hibbard}
\affiliation{Center for Astrophysics and Space Astronomy, Department of Astrophysical and Planetary Science, University of Colorado Boulder, CO 80309, USA}

\author[0000-0003-2196-6675]{David Rapetti}
\affiliation{NASA Ames Research Center, Moffett Field, CA 94035, USA}
\affiliation{Research Institute for Advanced Computer Science, Universities Space Research Association, Washington, DC 20024, USA}
\affiliation{Center for Astrophysics and Space Astronomy, Department of Astrophysical and Planetary Science, University of Colorado Boulder, CO 80309, USA}

\author[0000-0002-4468-2117]{Jack~O.~Burns}
\affiliation{Center for Astrophysics and Space Astronomy, Department of Astrophysical and Planetary Science, University of Colorado Boulder, CO 80309, USA}

\author[0000-0003-2560-8023]{Nivedita Mahesh}
\affiliation{California Institute of Technology, Pasadena, CA 91125, USA}

\author{Neil Bassett}
\affiliation{Center for Astrophysics and Space Astronomy, Department of Astrophysical and Planetary Science, University of Colorado Boulder, CO 80309, USA}

\email{Joshua.Hibbard@colorado.edu}
 
\begin{abstract}
Accurate detection of the cosmological 21-cm global signal requires galactic foreground models which can remove power over ~$10^6$. Although foreground and global signal models unavoidably exhibit overlap in their vector-spaces inducing bias error in the extracted signal, a second source of bias and error arises from inadequate foreground models, i.e. models which cannot fit spectra down to the noise level of the signal. We therefore test the level to which seven commonly employed foreground models---including nonlinear and linear forward-models, polynomials, and maximally-smooth polynomials---fit realistic simulated mock foreground spectra, as well as their dependence upon model inputs. The mock spectra are synthesized for an EDGES-like experiment and we compare all models' goodness-of-fit and preference using a Kolomogorov-Smirnov test of the noise-normalized residuals in order to compare models with differing, and sometimes indeterminable, degrees of freedom. For a single LST bin spectrum and p-value threshold of $p=0.05$, the nonlinear-forward model with 4 parameters is preferred ($p=0.99$), while the linear forward-model fits well with 6-7 parameters ($p=0.94,0.97$ respectively). The polynomials and maximally-smooth polynomials, like those employed by the EDGES and SARAS3 experiments, cannot produce good fits with 5 parameters for the experimental simulations in this work ($p<10^{-6}$). However, we find that polynomials with 6 parameters pass the KS-test ($p=0.4$), although a 9 parameter fit produces the highest p-value ($p\sim0.67$). When fitting multiple LST bins simultaneously, we find that the linear forward-model outperforms (a higher p-value) the nonlinear for 2, 5 and 10 LST bins. Importantly, the KS-test consistently identifies best-fit \textit{and} preferred models.
\end{abstract}


\section{Introduction}
\label{sec:intro}
Many windows into the early Universe---from the Cosmic Dawn to the Epoch of Reionization---have begun to open as low-frequency single antenna/radiometer and interferometer experiments, aimed at the elusive 21-cm emission line of neutral Hydrogen produced by the Inter-Galactic Medium \citep[IGM, see][]{madau_21-cm_1997,shaver_can_1999}, experience first light. Beginning with the Experiment to Detect the Global EoR Signature \citep[EDGES,][]{bowman_absorption_2018}, and followed by the Shaped Antenna Measurement of the Background Radio Spectrum \citep[SARAS3,][]{singh_detection_2021} and the Hydrogen Epoch of Reionization Array \citep[HERA,][]{the_hera_collaboration_hera_2022}, data now exist which have the potential to illuminate the nature of the first stars and galaxies \citep{furlanetto_cosmology_2006,pritchard_21-cm_2012,barkana_signs_2018, schauer_constraining_2019, munoz_impact_2022,furlanetto_bursty_2022}, measure small-scale structure formation and the tomography of Reionization \citep{munoz_probing_2020,mirocha_galaxy-free_2022}, and perhaps provide a new method for constraining dark matter \citep{furlanetto_effects_2006,barkana_possible_2018,fialkov_constraining_2018,schneider_constraining_2018,hibbard_constraining_2022}.

Other global and power-spectrum experiments, including the Large-Aperture Experiment to Detect the Dark Age \citep[LEDA,][]{price_design_2018,spinelli_antenna_2022}, Probing Radio Intensity at high-Z from Marion \citep[PRIZM,][]{philip_probing_2018}, Sonda Cosmol{\'o}gica de las Islas para la Detecci{\'o}n de Hidr{\'o}geno Neutro \citep[SCI-HI,][]{voytek_hydrogen_2015}, the Radio Experiment for the Analysis of Cosmic Hydrogen \citep[REACH,][]{de_lera_acedo_reach_2022}, the Mapper of the IGM Spin Temperature \citep[MIST,][]{monsalve_mapper_2023} and power spectrum experiments with data taken by the Murchison Widefield Array \citep[MWA,][]{yoshiura_new_2021}, the LOw-Frequency ARray \citep[LOFAR,][]{mertens_improved_2020}, and the Long Wavelength Array \citep[LWA,][]{eastwood_21_2019}, are under development or have released initial constraints.

The foremost obstacle to measuring the 21-cm signal, both global and power spectrum remains the galactic foreground, four to six orders of magnitude greater than the cosmological signal \citep{pritchard_21-cm_2012} and comprised primarily of synchrotron radiation generated by relativistic electrons spiraling in galactic magnetic fields \citep{condon_essential_2016}. Although the intrinsic foreground is well-modeled by spectrally-smooth power laws, when coupled to realistic, chromatic antenna beams the result is a spectral Gordian knot; that is, the frequency dependence of the beam couples spatial variations of the foreground sky as spectral variations into the received signal at the antenna. The ineluctable fact remains, seldom acknowledged, that models for the beam-weighted, galactic foreground must be able to fit the foreground component of the data \textit{down to the noise-level of the signal}. Residual total intensity beam-weighted foreground power can not only obscure the signal entirely, but lead to false signal reconstructions \citep{tauscher_formulating_2020,anstey_general_2021,rao_modeling_2017,bassett_lost_2021}, as can unmodelled intrinsic foreground polarization \citep{spinelli_antenna_2022}, or inadequately removed ionospheric ripples \citep{shen_quantifying_2021}; we focus on modelling the first in this paper. Any proper foreground model, then, ought to be able to fit realistic, mock foreground spectra down to the $\sim 20$ mK level or less (the experimental level reported by EDGES, with residual systematic error above the radiometer noise), constituting a requirement of nearly six orders of magnitude of foreground removal.

To be explicit, there are two main sources of systematic bias and additional uncertainties in the process of extracting the 21-cm signal. The first comes from the inevitable overlap of the foreground and the signal models, and cannot be mitigated wholly as their vector spaces are in general not orthogonal. However, 
 \textit{the second significant source comes from inadequate foreground models themselves.} If a foreground model alone cannot fit mock foreground-only spectra down to the noise-level of the signal, even with the addition of more parameters, then the residual level that is attained represents the ``limit'' of this foreground model, and one cannot hope to get better signal extractions than this level \textit{regardless of the overlap with the signal model}. This source of error has received less recent attention, if any, by the community. This paper is thus primarily concerned with testing foreground models to see when/if they reach this level of accuracy, given mock foreground-only spectra which mimic real observations in complexity.

For readers who are interested in modelling and/or mitigating the first source of bias and error, namely that of overlap, we refer them to e.g., \cite{tauscher_global_2020,anstey_use_2023}. Note that all foreground models used in this work have also been used successfully in joint-fits (i.e., fits with foreground and signal); see below where each model is referenced to its primary work.

Ideally, we would construct foreground models from all-sky temperature maps at the Cosmic Dawn and EoR frequencies (from $10 - 200$ MHz). However, due to the dearth of such maps, let alone the utter lack of any with error estimations, it is necessary to use one of two methods to build robust, full-coverage foreground models. The first method uses principal component analysis \citep[see][]{de_oliveira-costa_model_2008,zheng_improved_2017} to extract modes from published low-frequency sky temperature maps, such as those of \cite{haslam_408-mhz_1982,guzman_all-sky_2011} at 408 MHz and 45 MHz, respectively. These modes can then be used to fill in the missing frequencies through interpolation. In the second method, the temperature maps are parameterized by nonlinear emission and absorption physics, such as spectral indices of synchrotron emission or electron densities contributing to free-free absorption \citep{rao_gmoss_2016,anstey_general_2021}.

Both methods depend upon the spatial features revealed by sky temperature maps. Such maps require high numbers of spherical harmonics to fully characterize, and models reliant upon such decompositions can be prohibitively computationally expensive. The lack of errors in the temperature maps themselves also prohibits their direct use as a model in likelihoods where error characterization is essential. Thus it is necessary to use the published sky temperature maps as a baseline (a model input) for characterizing the spatial features of the low-frequency sky, and to then build flexible enough foreground models which can account for some amount of spatial variation or error in the underlying temperature maps.

The method of generating linear eigenmodes (via singular value decomposition, SVD), which characterize the most important modes of variation, from a matrix which consists of columns of spectra generated from a nonlinear, physically-motivated foreground model, as proposed in \cite{switzer_erasing_2014,vedantham_chromatic_2014, tauscher_global_2018}, lies somewhere between method one and two.

Finally, it is common in the literature to employ purely phenomenological foreground models which describe smoothly-varying spectral features, such as polynomials and maximally smooth functions. Although it was initially believed that these models would be sufficient to describe the intrinsic power-law foreground with relatively few ($\sim 3$) terms \citep{furlanetto_cosmology_2006,pritchard_21-cm_2012}, it became quickly clear that the large chromaticity of the beam distorted the smooth-structure sufficiently to vitiate fits with a small number of parameters \citep{harker_mcmc_2012,bernardi_bayesian_2016,mozdzen_limits_2016, sims_testing_2020}. Distortions from the ionosphere further increase the number of needed terms \citep{mozdzen_spectral_2019}. As such, increasing numbers of polynomial parameters are required to describe the resultant beam-weighted foreground spectra adequately, indelibly increasing the overlap (and thus extracted errors) with 21-cm signal models \citep[see e.g.,][]{bassett_ensuring_2021}.

We reemphasize that these two separate problems of modelling the foreground alone sufficiently, and of the foreground's inevitable overlap with signal models, are deeply coupled: residual power left from the former will almost certainly be absorbed by the latter and cause a biased signal extraction. To avoid this, more foreground parameters will probably be needed, which in turn, will increase the overlap with signal models, leading to larger signal errors. Furthermore, goodness-of-fit statistics may proclaim a fit at the noise level of the signal when in fact both foreground and signal have poor fits individually, but their unmodelled components combine to cancel in the residuals.

To understand the limits of foreground modelling alone, as well as which models are preferred when fitted to realistic mock foreground spectra, we present an analysis of seven foreground models commonly employed in the literature: two with spatial dependencies --one nonlinear based upon the Radio Experiment for the Analysis of Cosmic Hydrogen (REACH) model of \cite{anstey_general_2021} and one linear based upon eigenmodes and singular value decomposition from \cite{tauscher_global_2018}--, three common polynomial models in frequency $\nu$ or $\ln{\nu}$ \citep{bowman_absorption_2018}, and two maximally-smooth polynomial models \citep{rao_modeling_2017,bevins_maxsmooth_2021}.

We first construct mock foreground spectra from realistic spatial and spectral intrinsic foregrounds, and then convolve them using realistic beams, horizons, time-sampling, and noise levels. We test the ability of each model to fit the mock spectra down to the noise level, and compute several goodness-of-fit and model comparison statistics. We test not only how dependent the models are on their spatial inputs (where applicable), but how many terms are needed in each case for foreground-only fits to reach the signal noise level. Importantly, we also present a more robust method for comparing and choosing best-fits across all models (useful in the case where goodness-of-fit statistics are unreliable due to an inability to calculate degrees of freedom accurately, as in the case of nonlinear models such as those employed by REACH and SARAS), using normalized residuals in a Kolmogorov-Smirnov test.

This work is motivated in part by requirements for upcoming lunar-based telescopes that will soon begin global 21-cm observations from the uniquely radio-quiet, ionosphere-free, and environmentally stable lunar surface. These observations are enabled by NASA's Commercial Lunar Payload Services program that will deliver our radio instruments to the lunar farside including operations during lunar night \citep{burns_low_2021,burns_lunar_2021,bale_lusee_2023}.

We first construct our realistic foreground mock spectra in Section \ref{sec:mock-sim}, and then in Section \ref{sec-foreground-models} introduce the foreground models which we fit to the mock. Section \ref{sec-methodology} outlines the statistics used and our Bayesian analysis framework. We present the results of our fits in Section \ref{sec-results}, discuss the implications of our findings in Section \ref{sec-discussion}, and conclude in Section \ref{sec-conclusions}.

\section{Simulated Foreground Data (REACH Mock)}
\label{sec:mock-sim}
We begin by building a spatially and spectrally realistic, beam-weighted mock foreground. We refer to the unweighted galactic sky as the \textit{intrinsic} foreground, and the foreground measured by a radio antenna experiment as the \textit{beam-weighted} foreground. The former is generated by synchrotron emission from relativistic electrons spiraling in the galactic magnetic field. These electrons have power-law energy number density distributions, and thus produce power-law emission spectra \citep{condon_essential_2016}. The beam-weighted foreground is generated by coupling a radio beam to the intrinsic foreground.

\subsection{Intrinsic Foreground}
The intrinsic mock foreground we employ is the same as that of the REACH team \citep{anstey_general_2021,anstey_informing_2022} in their studies:
\begin{equation}
    T_{\text{fg}}(\Omega, \nu) = (T_{408}(\Omega) - T_{\text{CMB}}) \left(\frac{\nu}{408}\right)^{\beta_R(\Omega)} + T_{\text{CMB}}
    \label{eqn-int-foreground}
\end{equation}
where $T_{408}$ is the sky map generated at 408 MHz by the Global Sky Model (GSM) of \cite{de_oliveira-costa_model_2008}, $T_{\text{CMB}} = 2.725$ K is the temperature of the Cosmic Microwave Background (CMB), and the spectral index $\beta_R$ is given by
\begin{equation}
    \beta_R(\Omega) = \frac{\log\left( \frac{T_{230}(\Omega) - T_{\text{CMB}}}{T_{408}(\Omega) - T_{\text{CMB}}} \right)}{\log\left( \frac{230}{408}\right)},
    \label{eqn:reach-beta}
\end{equation}
where $T_{230}$ is the sky map generated at 230 MHz by the same GSM. We calculate all extrapolations and convolutions using the HEALPIX module of \cite{gorski_healpix_2005}.

\begin{figure}
    \centering
    \includegraphics[width=0.48\textwidth]{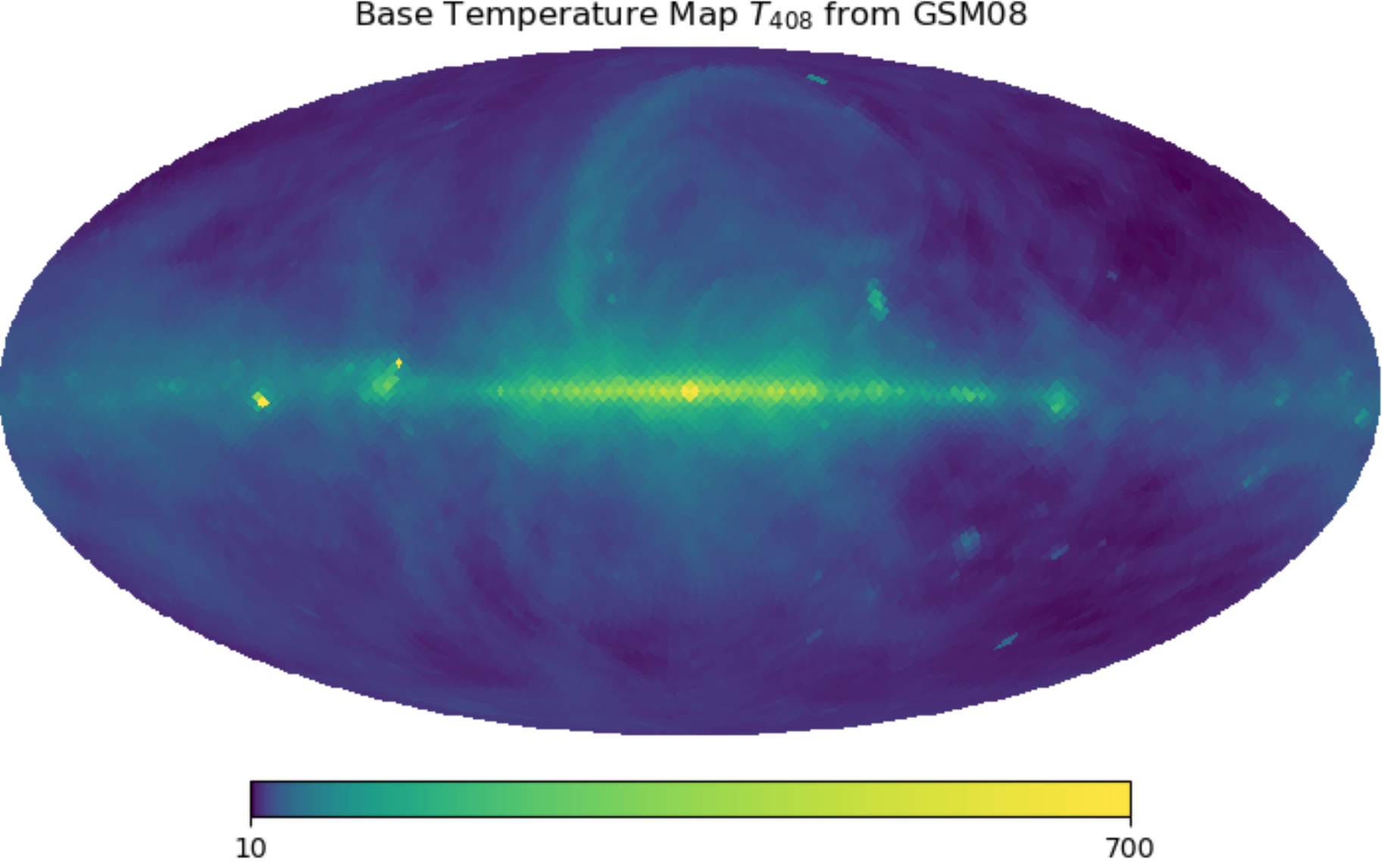}
    \includegraphics[width=0.48\textwidth]{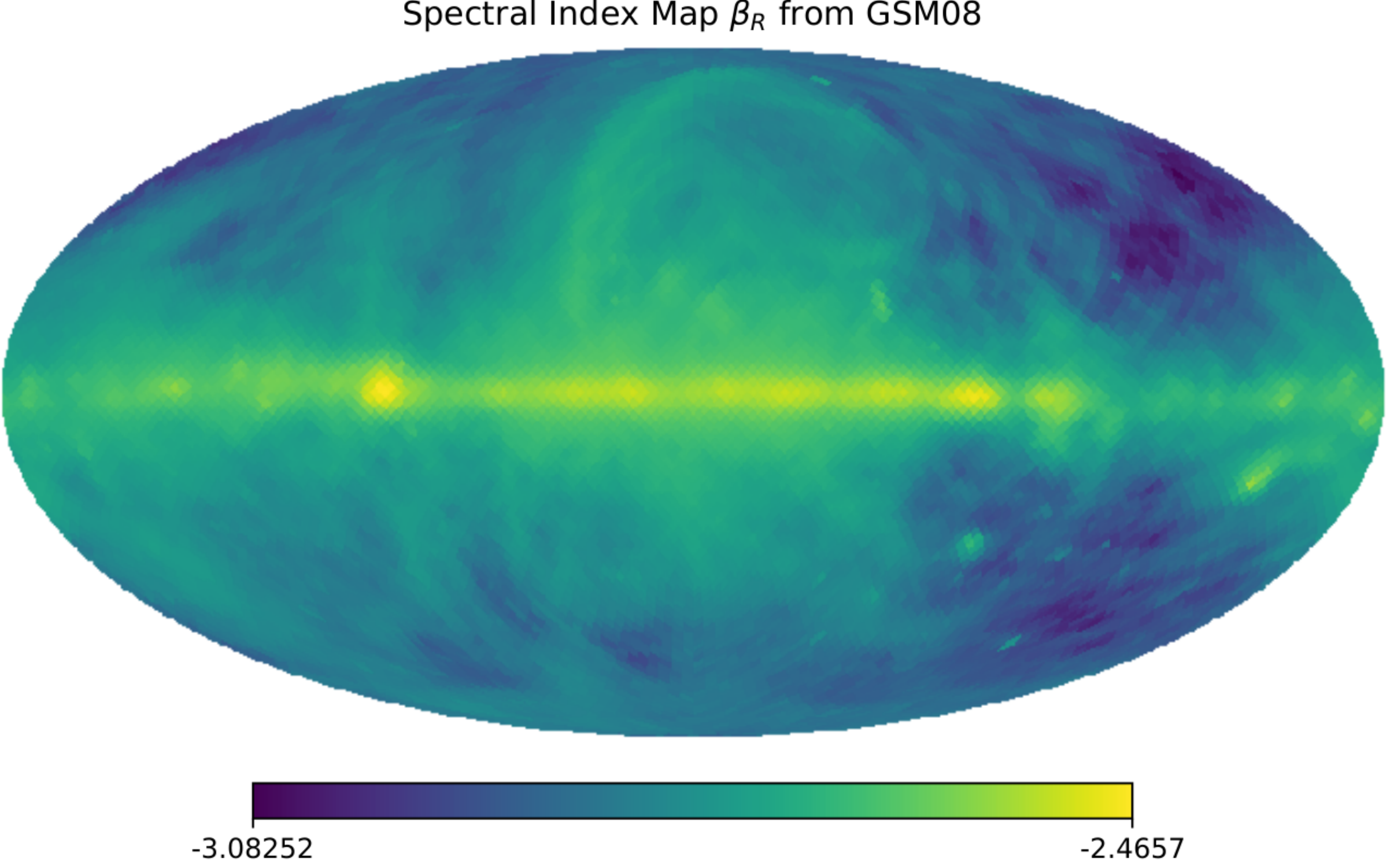}
    \caption{Base temperature map and spectral index map for the intrinsic foreground in the simulated data. The latter map is calculated by forming the natural logarithmic ratio of the two temperature maps generated by the GSM at 408 and 230 MHz, modulated by their reference frequency ratio.}
    \label{fig:mock-data-spectral-index}
\end{figure}

This intrinsic foreground mock exhibits both complicated spatial temperature and spectral distributions, and is thus a suitable representation of the inherent complexity of the true foreground. Our intrinsic foreground is shown in Figure \ref{fig:mock-data-spectral-index}, with the upper panel showing $T_{408}$ and the bottom panel $\beta_R$. 

However, in reality, it is likely that the true foreground deviates in spatial structure from 408 MHz down to 40 MHz, and contains small fluctuations in the spectral indices over the bandwidth of interest, known as spectral curvature \citep{mozdzen_improved_2017,mozdzen_spectral_2019,irfan_measurements_2021}. For this work, however, we neglect such intrinsic spectral variations, and assume that the spectral indices $\beta_R$ contain all spectral information, and $T_{408}$ all spatial temperature information.

\subsection{Beam Model, Horizon, and Observation Strategy}
To generate our beam-weighted foreground spectra from the intrinsic foreground mock, we assume a beam and observation strategy similar to the EDGES observatory, situated near the Murchison Radio Observatory (MRO) in Western Australia \citep{bowman_absorption_2018}. Therefore, we use a model of the antenna beam as generated by the computational electromagnetics software package FEKO\footnote{Altair FEKO - https://altairhyperworks.com/product/FEKO} for a blade antenna with an extended ground plane \citep[see][]{mahesh_validation_2021}.

\begin{figure}
    \centering
    \includegraphics[width=0.48\textwidth]{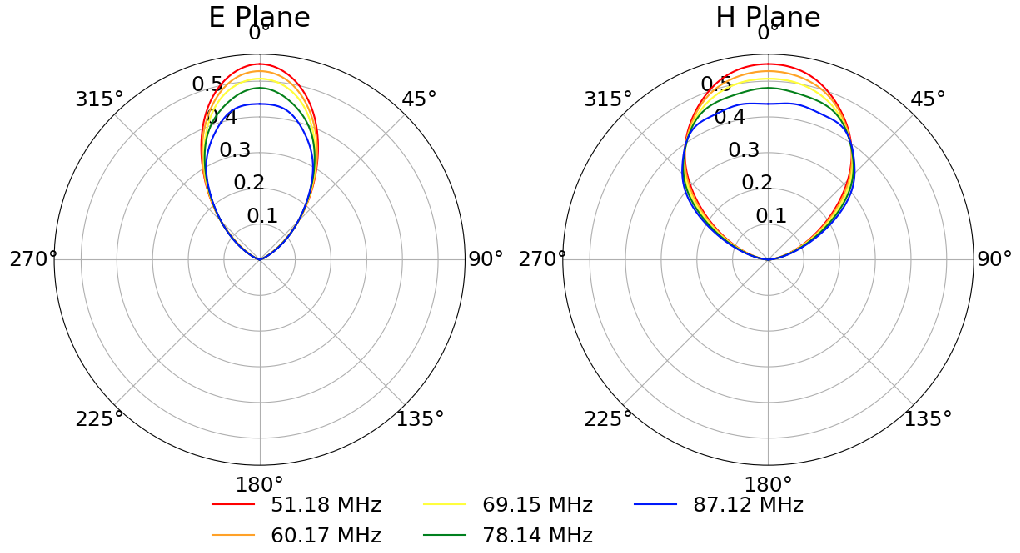}
    \caption{E and H Plane polar cuts of the EDGES antenna beam used to convolve the intrinsic foreground mock. The units shown are normalized, absolute units, such that the integral of the beam over the entire sky is unity at every frequency. The different colors represent the beam at various frequencies, as shown in the legend, from roughly 50 to 90 MHz.}
    \label{fig:input-beams}
\end{figure}

Figure \ref{fig:input-beams} shows the E and H Plane polar cuts of the EDGES antenna beams $B(\Omega)$ used to convolve the intrinsic foreground mock and generate the simulated spectra\footnote{These beam simulations have been provided courtesy of the EDGES collaboration.}.

Our horizon profile (the fraction of the $4 \pi$ steradian sky set to zero due to blocking by the ground and local topology, $H(\Omega)$) is generated by the SHAPES code from \cite{bassett_lost_2021} for the EDGES experiment's latitude and longitude. See the latter work for the calculation and visualization of the horizon profile\footnote{We normalize the beam over the entire sky in spherical coordinates before multiplying it by the horizon profile.}.

Lastly, note that the observing band for our mock spectra runs from 51 to 87 MHz, with $\Delta \nu \approx 0.4$ MHz channel spacing and a total number of channels $N_{ch} = 93$.

\subsection{Time-Sampling and Multiple LST Bins}
Most mock spectra of global experiments simulations assume continuous integration of the intrinsic foreground for some LST duration $\delta t$ \citep[see for example][]{tauscher_global_2020,anstey_general_2021}. For instance, one might assume integration for a duration of $\delta t = 6$ hours from LST 4 - 10 for mock observations. The effective foreground seen in this example is then a smeared version of the intrinsic foreground produced by the continuous integration and rotation of the sky from LST 4 to LST 10:
\begin{equation}
    T_{\text{eff}}(\Omega, \nu, t_i) = \frac{1}{2 \delta t} \int^{t_i + \delta t}_{t_i - \delta t} T_{\text{fg}}(\Omega, \nu, t) f_n(t) dt,
\end{equation}
where $t_i$ labels the middle of the LST bin, and $f_n(t)$ is a time-sampling function, described below. 

In practice, most experiments do not observe the sky continuously for hours or even minutes, but take spectra with short integration times of several seconds while the switch is in the antenna position \citep[see for instance, work by the EDGES collaboration][]{monsalve_calibration_2017,mozdzen_spectral_2019, monsalve_absolute_2021,mahesh_validation_2021}. As we seek to produce mock spectra as similar as possible to the actual EDGES low-band 2016-2017 run data, we follow the time-sampling of that experiment, which consists of taking $\delta t = 13$ second integrated spectra every 39 seconds. These spectra are then averaged together into LST bins according to the center of their LST time stamp $t_i$. This allows us to write down the effective foreground as seen by an antenna taking discrete time samples $t_i$ of length $\delta t$ as 
\begin{equation}
    \begin{split}
        T_{\text{eff,n}}(\Omega, \nu) = \frac{1}{2N_n\delta t} \sum^{N_n}_{i=1} \int^{t_i + \delta t}_{t_i - \delta t} T_{\text{fg}}(\Omega, \nu, t) dt,
    \end{split}
\end{equation}
where $N_n$ is the number of samples in each LST bin $n$.

Finally, we note that our method of LST-binning differs slightly from that employed in \cite{anstey_use_2023} by the REACH team. For this work, we separate the entire sky into equal-spaced bins based upon the desired number of total bins $n = [1,2,5,10]$, and then within each final LST bin we average together a number $N_n$ of higher resolution LST snapshots. Therefore, for example, if the sky is broken in to $n=2$ final bins, snapshots in LST ranges 1-12 would fall into the first final bin, and 13-24 into the second final bin. In the latter paper, typically one to three hours at a particular LST will then be split into higher-resolution bins of the order of minutes.

\subsection{Noise Estimate}
We simulate the noise for our mock spectra by assuming that the system temperature $T_{sys}$ is set by the sky temperature $T_{\text{fg}}$, equivalent to setting noise source temperatures to infinity and both load and receiver temperatures to zero. As we are comparing our mock spectra to those taken by the EDGES experiment during its 2016-2017 run, as seen in Hibbard et al. in prep, we calculate the noise in each LST bin as from an ideal radiometer, given by
\begin{equation}
    \sigma_{n}^2 = \frac{T^2_{n}(\nu)}{D_w},
    \label{eqn-noise-estimate}
\end{equation}
where the weighted, dynamic range $D_w = \Delta \nu \delta t W_n$, and $W_n$ are the weights (number of raw channels times number of raw spectra) in each LST bin, but are the same for every channel with a given bin. $T_n(\nu)$ refers to the final, convolved mock spectra in LST bin $n$ (see below). In practice, we estimate $D_w$ for use in our simulations by dividing the actual averaged, observed EDGES spectra in Hibbard et al. (in prep) by the final observed EDGES noise level as estimated from propagation of errors when averaging together raw spectra in time and frequency which has already undergone excision. This estimate is done solely to provide us with a comparable noise level to calculate for our mock spectra here, given a dynamic range comparable to that in the real experiment, and indeed this approximation is less than a factor of two off when used to estimate the noise level in the actual EDGES data. For real data such a simplification will not suffice, but for the simulations here we are only interested in generating reasonable noise levels given the mock spectra of the sky and comparable weighted, dynamic ranges. 

\subsection{Convolved Mock Spectra}
Given an intrinsic foreground mock $T_{\text{fg}}(\Omega,\nu,t_i)$, a horizon profile $H(\Omega)$, a beam model $B(\Omega, \nu)$, and time samples $t_i$ belonging to an LST bin $n$, our final convolved foreground mock is given by
\begin{equation}
    T_{n}(\nu) = \frac{1}{4\pi} \int_0^{4\pi} d\Omega B(\Omega, \nu) H(\Omega) T_{\text{eff,n}}(\Omega, \nu) + \hat{\sigma}_n
\end{equation}
where as throughout the paper, $N_n$ is the number of spectra in LST bin $n$, and $\hat{\sigma}_n$ is a noise realization drawn from a Gaussian distribution with zero mean and variance given by Equation \ref{eqn-noise-estimate}.

\begin{figure*}
    \centering
    \includegraphics[width=0.96\textwidth]{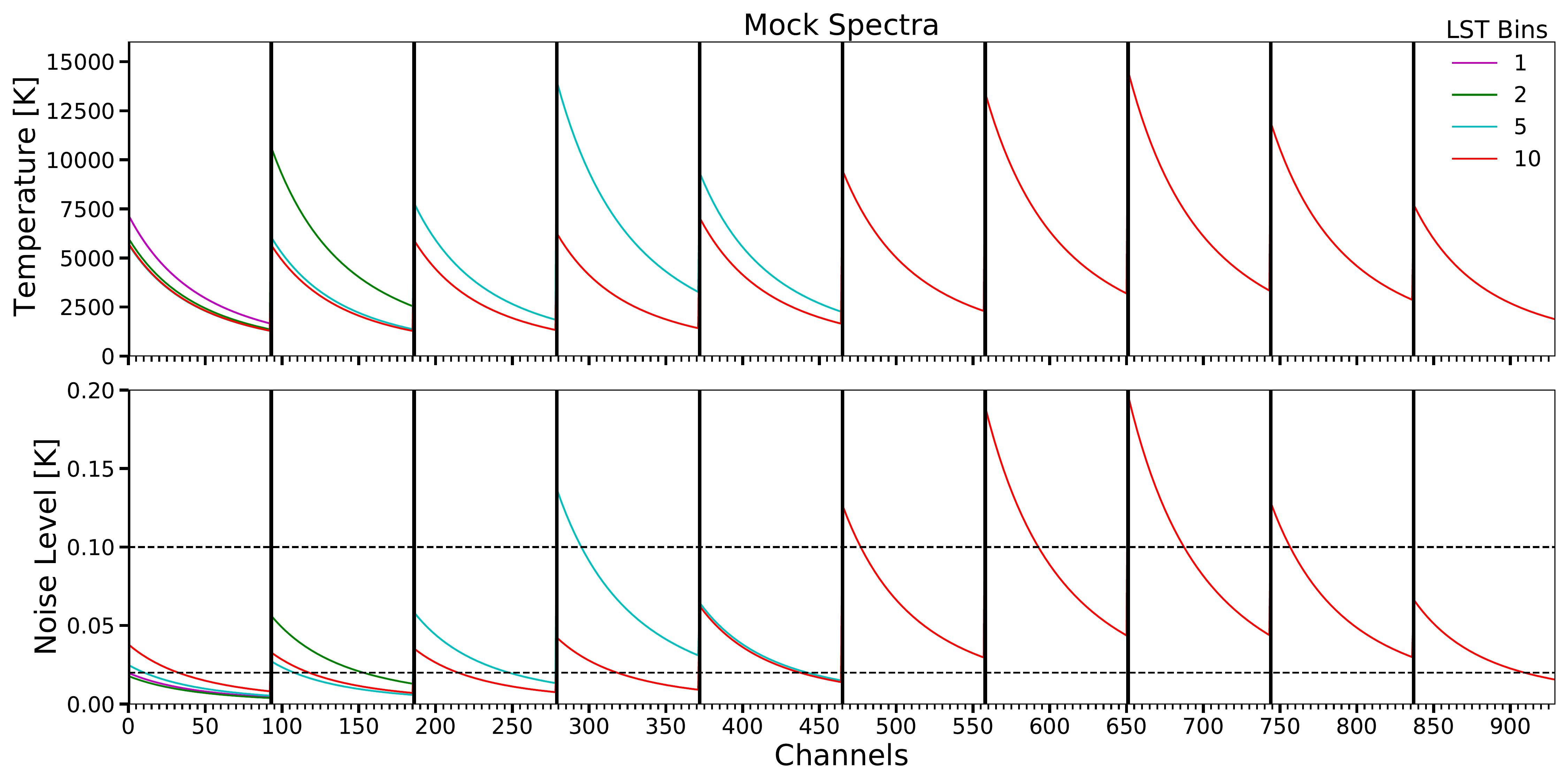}
    \caption{Simulated Mock Spectra (REACH Mock, top panel) and noise levels (bottom panel) plotted for various numbers of LST bins: 1 spectrum at LST 12 hr (magenta), 2 spectra at 6 and 18 hr (green), 5 spectra at 2.4, 7.2, 12, 16.8, 21.6 hr (cyan), and 10 spectra at 1.2, 3.6, 6, 8.4, 10.8, 13.2, 15.6, 18, 20.4, and 22.8 hr (red). The vertical black lines delineate each LST bin. We also plot the radiometer noise levels for the mock data (bottom panel), as well as flat 20 mK and 100 mK noise levels in black dashed lines for comparison. Black vertical lines delineate different spectra when the number of bins is greater than 1, and channels refers to a particular frequency in the bandwidth for a particular LST spectrum. Thus, channel 0 corresponds to 51 MHz for the first spectrum, while channel 93 refers to 51 MHz for the second spectrum, and so on.}
    \label{fig:mock_data}
\end{figure*}

In Figure \ref{fig:mock_data} we plot the final mock spectra (hereafter, the REACH mock) for each beam-weighted foreground of different LST bins, with the thick black lines delineating the individual spectra, concatenated together when $n > 1$. In the bottom panel of the figure, below each of the concatenated spectra, the corresponding noise levels used in the analysis are also shown. These $T_n$ represent our mock data that we shall now fit with galactic foreground models.

\section{Foreground Models}
\label{sec-foreground-models}

\begin{table*}[t]
    \setlength{\tabcolsep}{0.4em}
    \def\arraystretch{1.5}
    \centering
    \begin{tabular}{c|c|c|c|c|c|c|c}
        \hline
        \hline
        Model & Symbol & LSTs & Generating Equation & $\Theta$ &  $N_{\Theta}$ & Constraints & Priors \\
        \hline
        \hline
        \multirow{3}{*}{Nonlinear} & \multirow{3}{*}{$\boldsymbol{\mathcal{M}}_{\text{nl}}$} & \multirow{2}{*}{1,2,} & \multirow{3}{*}{\parbox{4.5cm}{\begin{equation}
                \sum_j^{N_r} K_j(\nu) A_j \left( \frac{\nu}{\nu_{0}}\right)^{\beta_j + \gamma_j \ln(\nu / \nu_0)} \label{eqn-nonlinear-ge}
            \end{equation}}} & $\beta_j$ & \multirow{2}{*}{$(1,2,3)$} & \multirow{3}{*}{--} & $\beta_j \sim U(-4.5, -2.0)$ \\ & & 5,10 & & $A_j$ & & & 
        $A_j \sim U(0.1, 10)$ \\ & & & $+ T_{\text{CMB}}$ & $\gamma_j$ & $ \times N_{r}$ & & 
        $\gamma_j \sim U(-0.1, 0.1)$ \\
        \hline
        \multirow{3}{*}{Linear} & \multirow{3}{*}{$\boldsymbol{\mathcal{M}}_{\text{lin}}$} & \multirow{2}{*}{1,2,} & \multirow{3}{*}{\parbox{4cm}{\begin{equation}
                \boldsymbol{F}_{\text{fg}}\boldsymbol{x}_{\text{fg}}
            \end{equation}}} & \multirow{3}{*}{$x^k_{\text{fg}}$} &  \multirow{3}{*}{$N_x$} & $\boldsymbol{F}_{\text{fg}} =$  cols$(\boldsymbol{U})$, & \multirow{3}{*}{$\boldsymbol{\pi_{\text{fg}}} \sim \mathcal{N}(\boldsymbol{\nu}; \boldsymbol{\Lambda})$} \\ 
        & & 5,10 & & & & $\boldsymbol{B}_{\text{fg}} = \boldsymbol{U} \boldsymbol{\Sigma} \boldsymbol{V}^T$ and & \\
        & & & & & &  col$(\boldsymbol{B}_{\text{fg}})_i=\boldsymbol{\mathcal{M}}_{\text{nl}}(\boldsymbol{\theta}_i)$ & \\
        \hline
        LinLogPoly & $\boldsymbol{\mathcal{M}_{\text{LinLogPoly}}}$ & 1 &  \parbox{4cm}{\begin{equation}
            \left( \frac{\nu}{\nu_0} \right)^{-2.5} \sum_{k=1}^{N_{py}} a_k \left[ \ln{\frac{\nu}{\nu_0}} \right]^k
            \label{eqn-lin-log-poly}
        \end{equation}} & $a_k$ & $N_{py}$ & -- & $\boldsymbol{\pi}_{\text{poly}} \sim \mathcal{N}(0;\sigma^2_{\text{poly}})$ \\
        \hline
        LinPoly & $\boldsymbol{\mathcal{M}_{\text{LinPoly}}}$ & 1 & \parbox{4cm}{\begin{equation}
                \left( \frac{\nu}{\nu_0} \right)^{-2.5} \sum_{k=1}^{N_{py}} a_k \left( \frac{\nu}{\nu_0} \right)^k
                \label{eqn-lin-poly}
            \end{equation}} & $a_k$ & $N_{py}$ & -- & $\boldsymbol{\pi}_{\text{poly}} \sim \mathcal{N}(0;\sigma^2_{\text{poly}})$ \\
        \hline
        \multirow{3}{*}{LinPhys} & \multirow{3}{*}{$\boldsymbol{\mathcal{M}_{\text{LinPhys}}}$} & \multirow{3}{*}{1} & \multirow{3}{*}{\parbox{4.5cm}{\begin{equation}
                 \left( \frac{\nu}{\nu_0} \right)^{-2.5} \sum^2_{k=1} a_k \left( \ln{\frac{\nu}{\nu_0}} \right)^k \label{eqn-lin-phys-poly}
            \end{equation}}} & \multirow{3}{*}{$a_k$} & \multirow{3}{*}{$5$} & \multirow{3}{*}{--} & \multirow{3}{*}{$\boldsymbol{\pi}_{\text{poly}} \sim \mathcal{N}(0;\sigma^2_{\text{poly}})$} \\
        & & & & & & & \\
        & & & \parbox{4.5cm}{$+ a_3 \left( \frac{\nu}{\nu_0} \right)^{-4.5} + a_4 \left( \frac{\nu}{\nu_0} \right)^{-2}$} & & & & \\
        \hline
        MSF & \multirow{2}{*}{$\boldsymbol{\mathcal{M}}_{MDP}$} & \multirow{2}{*}{1} & \multirow{2}{*}{\parbox{4cm}{\begin{equation}
            \sum_{k=1}^{N_{MSF}} a_k (\nu - \nu_0)^k \label{eqn-mdp}
        \end{equation}}} & \multirow{2}{*}{$a_k$} & \multirow{2}{*}{$N_{MSF}$} & \multirow{2}{*}{$\boldsymbol{G}\boldsymbol{a} \leq \boldsymbol{0}$} & \multirow{2}{*}{--} \\
        DiffPoly & & & & & & & \\
        \hline
        MSF & \multirow{2}{*}{$\boldsymbol{\mathcal{M}}_{MLLP}$} & \multirow{2}{*}{1} & \multirow{2}{*}{\parbox{4cm}{\begin{equation}
            10^{\sum_{k=1}^{N_{MSF}} a_k (\log_{10}{\nu})^k} \label{eqn-mllp}
        \end{equation}}} & \multirow{2}{*}{$a_k$} & \multirow{2}{*}{$N_{MSF}$} & \multirow{2}{*}{$\boldsymbol{G}\boldsymbol{a} \leq \boldsymbol{0}$} & \multirow{2}{*}{--} \\
        LogLogPoly & & & & & & & \\
        \hline
    \end{tabular}
    \caption{All galactic foreground models considered in this work, where each row denotes a different model. We fit each model to the REACH mock for a given set of LST bins, parameters, and model inputs. The Nonlinear model is the same as that employed by the REACH collaboration; the linear model generates eigenspectra from the nonlinear model via Singular Value Decomposition; the next three are all polynomial models; the last two are maximally-smooth polynomial models. The generating equation gives the model spectra for a given set of parameter values and model inputs. Parameter labels, the number of parameters, constrain equations, and priors are shown in the last four columns.}
    \label{tab:fg-models}
\end{table*}

We fit the REACH mock with four categories of foreground models: nonlinear forward-model, linear forward-model, polynomials (linear), and maximally smooth polynomials (nonlinear). The first two model categories are \textit{forward-models} in the sense that they rely upon spatial and spectral knowledge of model inputs such as the intrinsic foreground and beam, whereas the polynomial and maximally smooth models are purely spectral, phenomenological models. For brevity, we shall simply refer to the first two models as \textit{nonlinear} and \textit{linear}, respectively. In this work we seek to test not only how well each foreground model category can fit the REACH mock for a given set and number of parameters, but also how reliant upon the model inputs the first two categories are. In particular, we will test the robustness of assumptions about the spatial distribution of the base temperature map (BM) and of the spectral index patch map (PM; see below) which extrapolates the BM to the desired frequency band.

Table \ref{tab:fg-models} summarizes the seven foreground models tested in this work, including the nonlinear, linear, 3 polynomial, and 2 maximally-smooth polynomial models. They are delineated by model symbol, the number of LST bins to which they are applied (and consequently, the number of LST bins in the REACH mock they fit), the equation which generates the model spectra, the parameters in the model $\Theta$, the number of parameters in a given fit $N_{\Theta}$, the constraints upon the generating equation, and the parameter priors, if any. Only one model has a fixed number of parameters, the Linear Physical Polynomial $\boldsymbol{\mathcal{M}}_{LinPhys}$, with $N_{py} = 5$. 

The rest of this section will be devoted to further illuminating the assumptions made and calculations involved for each model. Readers familiar with the galactic foreground model literature and Bayesian analysis may wish to skip directly to Section \ref{sec-results} for the results of fitting each model to the REACH mock using standard Bayesian analysis.

\subsection{Nonlinear (Physical) Foreground Model}
\begin{figure}
    \centering
    \includegraphics[width=0.48\textwidth]{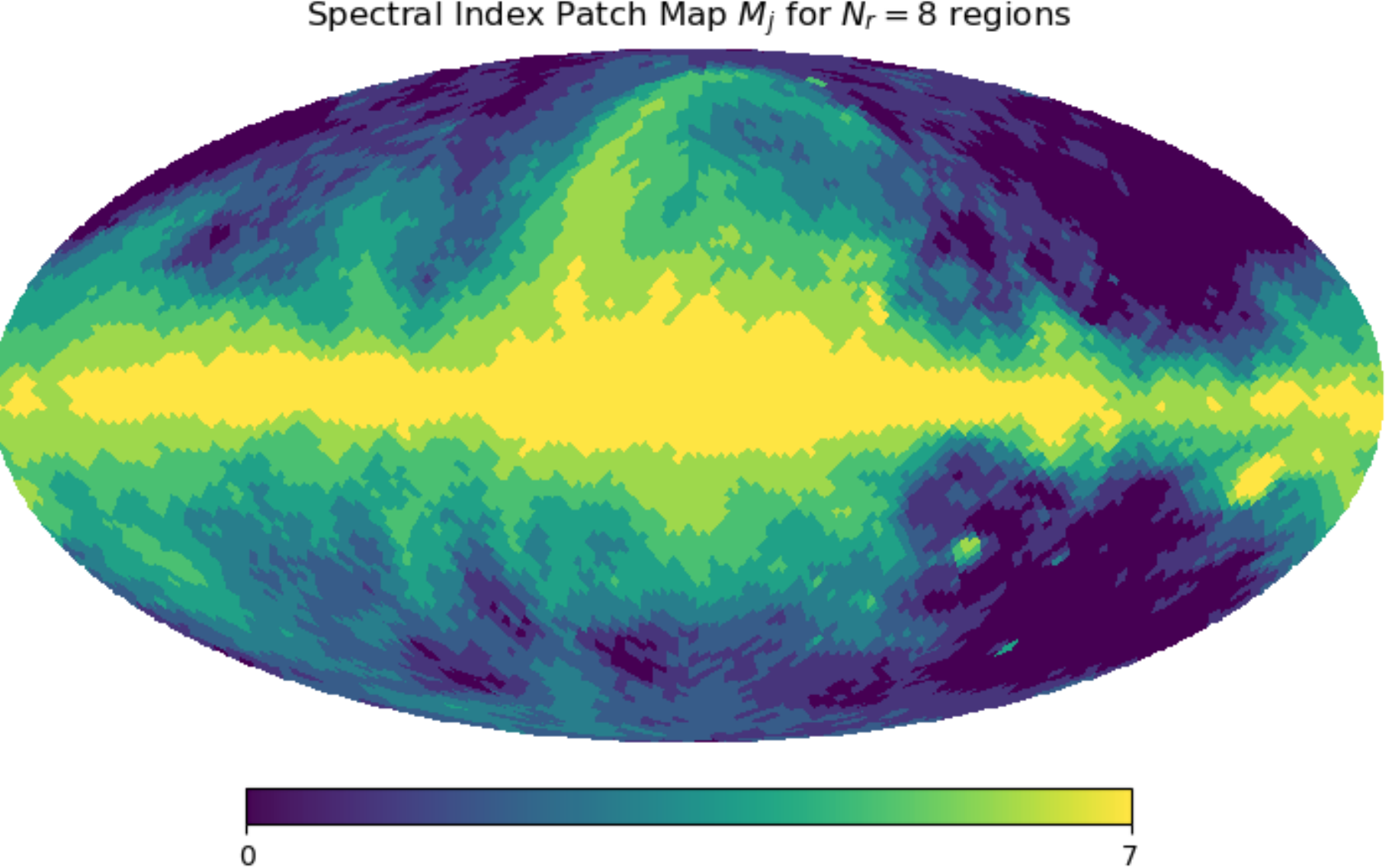}
    \caption{Approximation of the intrinsic spectral index map by dividing the sky into $N_r$ equal-percentile bin regions. Each region $j$ is then assigned parameters which are constant throughout the region. This particular map is generated from the spectral index map used in the intrinsic mock foreground, $\beta_R$.}
    \label{fig:patch-map-model}
\end{figure}

Based on work from the REACH collaboration \citep{anstey_general_2021,anstey_informing_2022,pagano_general_2022,de_lera_acedo_reach_2022}, we assume that $T_{\text{fg}}$ in Equation \ref{eqn-int-foreground} is well-modelled to the chosen noise level by approximating its spectral index map $\beta_R$ in Equation \ref{eqn:reach-beta} as a set of $N_r$ regions or ``patches'', each with roughly the same number of pixels. These regions are chosen by sorting the spectral index map pixels into $N_r$ equal-percentile bins, where each bin defines a region on the sky. For instance, a sky consisting of 4 regions would have a $0 - 25$ percentile bin, or region, a $26-50$ percentile region, and so on. Figure \ref{fig:patch-map-model} shows an example of this spectral index patch map when $N_r = 8$ regions. Each region $j$ is then assigned a constant spectral index $\beta_j$, magnitude $A_j$, and spectral curvature $\gamma_j$\footnote{Note that a perfect representation of Equation \ref{eqn:reach-beta} would require an $N_r$ in this model equal to the number of pixels in the spectral index input map.}. The curvature parameters allow for spectral structure poorly captured by the spectral index to be accounted for, as well as intrinsic curvature in the foreground (neglected here), while the magnitude parameters allow for a rescaling of the temperature base map\footnote{Amplitude scale factors were independently presented in a recent paper using the REACH pipeline and shown to be important to account for measurement errors in the base map \citep{pagano_general_2022}. In that work, the regions of these factors are designed to be independent of the spectral regions.}.

The above nonlinear parameters are shown in Equation \ref{eqn-nonlinear-ge} of Table \ref{tab:fg-models}, and in its corresponding $\Theta$ column. The so-called \textit{chromaticity functions} $K_j(\nu)$ in that equation contain all of the information concerning the observation strategy, time-sampling, beam, horizon, temperature base map (BM) and spectral index patch map (PM). They are parametrized by region $j$ and calculated via
\begin{equation}
    \begin{split}
        K_j(\nu) = \frac{1}{4\pi} \int_0^{4\pi} d\Omega B(\Omega, \nu) H(\Omega) M_j(\Omega) \\ \times \frac{1}{2N_n\delta t} \sum^{N_n}_{i=1} \int^{t_i + \delta t}_{t_i - \delta t} (T_{0}(\Omega, \nu, t) - T_{\text{CMB}}) dt.
    \end{split}
    \label{eqn-chromaticity-functions}
\end{equation}
where the different PM regions are formed by the masks $M_j$, which are unity for each pixel within a region $j$, and zero otherwise. $M_j$ is calculated from $\beta_R$ (the same as the REACH mock) for the primary case of the nonlinear model, although see below where we also test for a case when $M_j$ is not derived from $\beta_R$. This set of masks is one of the two model inputs required to build the nonlinear model. The other model input, the base map (BM) $T_0$ at reference frequency $\nu_0$, is likewise set to $T_{408}$ for the primary, or ideal, case. We also test the case in which $T_{0} \neq T_{408}$. For $N_r$ sky regions, there can be up to $N_{nl} = 3N_r$ parameters for this model, depending on whether we include magnitude $A_j$ and spectral curvature $\gamma_j$ parameters. When these parameters are not included in a fit, we set their values to $1$ and $0$, respectively.

\subsection{Linear Foreground Model}
The linear foreground model we employ in this work is, in essence, a \textit{linearized} version of the nonlinear model given above. For a given set of nonlinear model inputs, such as the number of regions $N_r$, the BM and PM, and the nonlinear parameters $\boldsymbol{\theta}$, we simulate ten thousand nonlinear model spectra by computing $\boldsymbol{b}_{draw,i} = \boldsymbol{\mathcal{M}}_{\text{nl}}(\boldsymbol{\theta}_{draw,i})$ for each nonlinear parameter draw $\boldsymbol{\theta}_{draw,i}$ labelled by $i$. We generate the latter by sampling from uniform prior ranges similar to those used for the nonlinear model priors, albeit slightly truncated: $\beta_j \sim U(-3.5,-2.0)$, $A_j \sim U(0,3)$, $\gamma_j \sim U(-0.1,0.1)$. Each simulated nonlinear spectrum $\boldsymbol{b}_{draw,i}$ is then stacked horizontally into a matrix denoted $\boldsymbol{B}_{\text{fg}}$ as one of its columns. This matrix thus has dimensions $N_{ch} \times N_{sp}$, where $N_{ch}$ is the number of channels in each spectra (number of frequencies times the number of LST bins concatenated together) and $N_{sp} = 10,000$ is the number of simulated spectra. We then compute the eigenmodes of this matrix of spectra, hereafter referred to as \textit{eigenspectra}, via Singular Value Decomposition (SVD) -- for details on SVD, we refer the reader to our \texttt{pylinex}\footnote{\url{https://github.com/CU-NESS/pylinex}} pipeline series \citep{tauscher_global_2018, rapetti_global_2020, tauscher_global_2020,tauscher_global_2021}, as well as to recent work by the REACH collaboration building on our SVD techniques \citep{saxena_sky-averaged_2022}. The SVD eigenspectra, given mathematically as the columns of $\boldsymbol{F}_{\text{fg}}$, then represent the modes of our linear model and provide the best fit to every spectrum in the matrix $\boldsymbol{B}_{\text{fg}}$. In this case, best fit means that the eigenspectra derived from SVD minimize the total root-mean-square (RMS) of all residuals for fits to every spectrum in the matrix. 

Theoretically we would need a number of eigenspectra equal to the rank of the matrix to span its null space, but in practice we truncate the number of vectors when we fit the desired mock data to the input noise level. This typically requires $5 - 50$ eigenspectra for one or two LST bins, but can require up to $150$ eigenspectra for 10 LST bins. The parameters in this model are labelled as $\Theta = [\boldsymbol{x}_{\text{fg}}]$, which is a parameter vector of length $N_{x}$ equal to the number of eigenspectra, or columns in $\boldsymbol{F}_{\text{fg}}$.

To generate priors for the linear parameters $\boldsymbol{x_{\text{fg}}}$, we first fit every spectrum in $\boldsymbol{B}_{\text{fg}}$ with its own eigenspectra. We then calculate the mean and covariance of each linear model parameter $x_{\text{fg}}^i$ across all fits, and use these values to define a multivariate Gaussian prior distribution for the linear model parameters with mean $\boldsymbol{\nu}$ and diagonal covariance $\boldsymbol{\Lambda}$ \citep[see also][for details on this calculation]{bassett_ensuring_2021,tauscher_global_2020}. Using these priors assumes that the magnitudes of the spectra in the simulated matrix $\boldsymbol{B}_{\text{fg}}$ well-describe the magnitudes of the underlying convolved foreground, hence further relying on the validity of the foreground model.

\subsection{Polynomial Foreground Models}
As there is no obvious way to generalize the polynomial foreground models to fit multiple LST bins \textit{simultaneously}, these models are limited to the case of 1 LST bin. We focus on three polynomial models commonly employed in the literature \citep[see e.g.,][]{bowman_absorption_2018,singh_detection_2021,bevins_astrophysical_2022}: linear-logarithmic (LinLogPoly), linear (LinPoly), and linearized physical (LinPhys). 

\subsubsection{Traditional Polynomial Models}
\label{sec:trapoly}

The LinLogPoly model consists of powers of natural logarithms, while the LinPoly model consists of mere polynomials in $(\nu / \nu_0)$. Incidentally, the former model is a consequence of Taylor-expanding a foreground consisting of many smooth power-laws added together \citep{hibbard_modelling_2020}, though including a chromatic beam factor to each power-law disrupts this smoothness and thus this approximation\footnote{For an alternative form of the LinLogPoly model, see the following memo by the EDGES collaboration: \url{https://loco.lab.asu.edu/loco-memos/edges_reports/report122.pdf}. This form factors out a power-law before Taylor-expanding, as opposed to what is done in \citep{hibbard_modelling_2020}. We retain our version of the polynomial model for comparison to other works which use the same form.}. The LinPhys model is supposed to account for both the galactic foreground and ionospheric distortion and is similarly derived from a Taylor-expansion of a physically-motivated, nonlinear, power-law based foreground model \citep{bowman_absorption_2018}. 

In each polynomial model, the $\Theta = [a_k]$ are the (linear) polynomial parameters, and $N_{py}$ refers to the number of polynomial parameters in each fit. We select broad, Gaussian priors with zero mean and variance given by $\sigma^2_{\text{poly}} = (3000)^2$.

\subsubsection{Maximally Smooth Polynomial Models}
\label{sec:smoothpoly}

Maximally smooth polynomials are polynomial models with the added constraint that their higher-order derivatives have no zero-crossing points. That is, for the two models considered here, we require
\begin{equation}
    \begin{split}
        \frac{d^m \boldsymbol{\mathcal{M}}_{MDP}}{d\nu^m} \leq 0 \\
        \frac{d^m \log_{10} (\boldsymbol{\mathcal{M}}_{MLLP})}{d \log_{10}( \nu)^m} \leq 0
    \end{split}   
\end{equation}
where $\boldsymbol{\mathcal{M}}_{MDP}$ refers to the maximally smooth Difference Polynomial in Equation \ref{eqn-mdp}, and $\boldsymbol{\mathcal{M}}_{MLLP}$ refers to the maximally smooth Log Log Polynomial in Equation \ref{eqn-mllp}. The order of the derivative is labelled by $m$, and it is typically required that $m \geq 2$. Such constraints can be gathered into a matrix equation \citep{bevins_maxsmooth_2021} given by $\boldsymbol{G}\boldsymbol{a} \leq \boldsymbol{0}$, where $\boldsymbol{G}$ denotes the matrix of zeroes and derivatives of all desired orders, and $\boldsymbol{a}$ is a vector of the maximally smooth polynomial parameters, with length $N_{MSF}$.

These models were chosen as they appear best-suited to fit mock foreground spectra and are commonly employed in maximally smooth fits, as seen in \cite{bevins_maxsmooth_2021} and \cite{rao_modeling_2017}. These derivative constraints ensure that the polynomials fit only smooth foregrounds characterized by spectral indices and curvatures, and not global signals, which likely have multiple turning points and would thus be ill-fit by such smooth models \citep[see again][]{rao_modeling_2017,bevins_maxsmooth_2021,singh_detection_2021}. However, it should be noted that in the literature maximally-smooth polynomials have largely been used to fit spectra with achromatic beams, spectrally constant foregrounds, or noise levels on the order of $\sim 500$ mK, much too high for most predicted global signal trough amplitudes. See the latter references for examples.

\section{Methodology}
\label{sec-methodology}

\subsection{Bayesian Analysis}
Given a data vector $\boldsymbol{T}_{\text{n}}$ and a model $\boldsymbol{\mathcal{M}}_i$ with (linear or nonlinear) parameters $\boldsymbol{\Theta}$ from a set of possible models $[\boldsymbol{\mathcal{M}}_i]$, the posterior distribution, or the probability of the parameters given the data, is calculated from Bayes' theorem:
\begin{equation}
    \begin{aligned}P(\boldsymbol{\mathcal{M}}_i(\boldsymbol{\Theta})|\boldsymbol{T}_{\text{n}}) &= \frac{P(\boldsymbol{T}_{\text{n}}|\boldsymbol{\mathcal{M}}_i(\boldsymbol{\Theta})) P(\boldsymbol{\Theta})}{P(\boldsymbol{T}_{\text{n}})} \\ &= \frac{\mathcal{L} \pi}{Z}
    \end{aligned}
\end{equation}
where $\mathcal{L}$ is the likelihood, or the probability of the data given the model and parameters; $\pi$ is the prior, or probability of the parameters for a given model; and the normalizing factor $Z$ is the evidence, given by:
\begin{equation}
    Z = \int d\theta^{N_{\Theta}} \mathcal{L}(\boldsymbol{\Theta}) \pi(\boldsymbol{\Theta}),
    \label{eqn-evidence}
\end{equation}
which is the marginal likelihood over the prior volume. If the prior volume is uniform, it is a measure of the likelihood volume.

Assuming Gaussian-distributed noise, the natural logarithm of the likelihood is proportional to
\begin{equation}
    \ln{\boldsymbol{\mathcal{L}}} \propto (\boldsymbol{T}_{\text{n}} - \boldsymbol{\mathcal{M}}_{\text{i}})^T \boldsymbol{C}^{-1}(\boldsymbol{T}_{\text{n}} - \boldsymbol{\mathcal{M}}_{\text{i}}),
\end{equation}
where $\boldsymbol{C}$ is the noise covariance matrix of the data, which is typically written as a symmetric, diagonal matrix by assuming that all cross-correlations between channels are zero. In our simulations, the diagonal elements of this matrix are given by the noise level estimates in Equation \ref{eqn-noise-estimate}. 

Bayesian analysis then proceeds by seeking the model parameters which maximize the posterior distribution. If the priors are uniform, the parameters which maximize the likelihood are the same as the parameters which maximize the posterior. The former are referred to as the \textit{maximum likelihood estimate} parameters (MLE), while the latter are the \textit{maximum a posterior} parameters (MAP).

For the nonlinear model, we calculate the MAP parameters $\boldsymbol{\theta}_{\text{MAP}}$ and the evidence $\ln{Z}$ using $\texttt{polychord}$ with the default number of live points, repeats, and evidence tolerance \citep{handley_polychord_2015}. We run our nonlinear models on the BLANCA CURC condo-cluster at the University of Colorado. In general, these nonlinear fits must be performed individually for a given number of spectral regions $N_r$, and nonlinear parameters. 

If the model is linear, the MAP parameters can be calculated analytically:
\begin{equation}
    \begin{aligned}
        \boldsymbol{x}_{\text{MAP}} = \boldsymbol{S}_{MAP}(\boldsymbol{F}_{\text{fg}}^T \boldsymbol{C}^{-1}\boldsymbol{T}_{\text{n}} + \boldsymbol{\Lambda}^{-1} \boldsymbol{\nu}) \\
        \boldsymbol{S}_{\text{MAP}} = (\boldsymbol{F}_{\text{fg}}^T \boldsymbol{C}^{-1} \boldsymbol{F}_{\text{fg}} + \boldsymbol{\Lambda}^{-1})^{-1}.
    \end{aligned}
    \label{eqn-linear-map}
\end{equation}
where $\boldsymbol{S}_{\text{MAP}}$ is the MAP parameter covariance.

If the model is linear and the priors are Gaussian-distributed, we can also calculate the evidence analytically:
\begin{equation}
    Z = \sqrt{\frac{|\boldsymbol{S_{\text{MAP}}}|}{|\boldsymbol{\Lambda}|}} \exp{\frac{\boldsymbol{\mu}^T \boldsymbol{S_{\text{MAP}}} \boldsymbol{\mu} - \boldsymbol{T}_n^T \boldsymbol{C}^{-1} \boldsymbol{T}_n - \boldsymbol{\nu}^T \boldsymbol{\Lambda} \boldsymbol{\nu}}{2}}
    \label{eqn=anal-lin-ev}
\end{equation}
where $\boldsymbol{\mu} = \boldsymbol{F}_{\text{fg}}^T\boldsymbol{C}^{-1}\boldsymbol{T}_n + \boldsymbol{\Lambda}^{-1} \boldsymbol{\nu}$. See Appendix \ref{app-evidence} for a derivation of the above result.

While the number of parameters in the nonlinear model is a function of $N_r$, the number of parameters for the linear model, $N_x$, is determined via a grid search. Given a particular set of eigenspectra $\boldsymbol{F}_{\text{fg}}$, we compute the evidence for each possible number $N$ of linear parameters in a one-dimensional grid with $N \leq 100$ for 1 and 2 LST bins, and $N \leq 150$ for 5 and 10 LST bins. Moreover, we define $N_x$ to be the number of parameters needed to fit the spectra in the matrix $\boldsymbol{B}_{\text{fg}}$ down to the noise level, as determined by the $N$ with the highest evidence from the grid search, and is typically $N_x = 7$ for the 1 and 2 LST bin cases, and $N_x = 15$ or more for the 5 and 10 LST bin cases. Note that this grid search is performed for every different model input (BM, PM, and $N_r$). $N_x$ is then employed as the number of parameters used in the linear fit.

For the traditional polynomial models in Section~\ref{sec:trapoly}, we obtain the best-fit parameters $\boldsymbol{a}_{\text{MAP}}$ using Equation \ref{eqn-linear-map} with $\boldsymbol{F}_{\text{fg}}$ replaced by the polynomial modes defined in Equations \ref{eqn-lin-log-poly} - \ref{eqn-lin-phys-poly}.

For the maximally smooth polynomial models in Section~\ref{sec:smoothpoly}, we use \texttt{maxsmooth} from \cite{bevins_maxsmooth_2021} to compute the MAP parameters.

\subsection{Goodness of Fit}
While the evidence can be used to determine a preferred model for a given data set, it does not quantify the goodness-of-fit of the model to the data set. Instead, the reduced chi-squared statistic is often employed for this purpose:
\begin{equation}
    \chi^2_{red} = \frac{\boldsymbol{\delta}^T\boldsymbol{C}^{-1}\boldsymbol{\delta}}{N_{DOF}}
\end{equation}
where $\boldsymbol{\delta} = \boldsymbol{T}_n - \boldsymbol{\mathcal{M}}(\boldsymbol{\Theta}_{\text{MAP}})$ are the residuals of the best-fit, and $N_{DOF}$ is the number of degrees of freedom of the fit. If the model is linear and there are no priors, the expected value and variance of the $\chi^2_{red}$ statistic are $E[\chi^2_{red}] = 1$ and $V[\chi^2_{red}] = 2/N_{DOF}$, allowing for a calculation of the number of standard deviations $\sigma$ away from the expected value that a particular model fit produces. Robustly, the degrees of freedom for a linear model without priors is equal to the rank of the basis matrix, in this case the rank of $\boldsymbol{F}_{\text{fg}}$, \textit{if the vectors in the basis are linearly independent} \citep{andrae_dos_2010}. SVD produces orthogonal vectors, which satisfy this requirement; otherwise, $N_d - N_{\Theta} \leq N_{DOF} \leq N_d - 1$, where $N_d$ is the number of channels in the data vector. Therefore, for the linear models here, $N_{DOF} = N_d - N_{\Theta}$. When priors are included in linear models, $E[\chi^2_{red}] \geq 1$, and the expectation value can be calculated analytically.

\subsubsection{Challenges of Reduced Chi-Squared}

The trouble emerges when the model is nonlinear. In that case, not only is it in general impossible to calculate $E[\chi^2_{red}]$ when priors are included, but the number of degrees of freedom lies somewhere between zero and $N_d - 1$, may change during the fitting procedure, and subsequently cannot be estimated using the same equation for the linear model. Rigorously, this is because in the case of a linear model it is possible to write down a basis matrix with linearly independent vectors ($\boldsymbol{F}_{\text{fg}}$, for example). The rank of this basis, or design matrix, is equal to the degrees of freedom; \citep[see e.g.][]{ye_measuring_1998,hastie_elements_2009}. This means that in general, $E[\chi^2_{red}] \neq 1$ for nonlinear models, and without a method to estimate $E[\chi^2_{red}]$, determining the quality of fit based upon how close $\chi^2_{red}$ is to unity is grossly inadequate for nonlinear models, especially when considering the dynamic ranges involved in 21-cm signal extractions.

\subsubsection{KS-Test of Normalized Residuals}
An alternative statistic that can be used to characterize both goodness-of-fit \textit{and} model comparison is the set of noise-weighted residuals in a Kolmogorov-Smirnov (KS) test \citep{kolmogorov_a_n_sulla_1933,smirnov_table_1948}. By definition for our Gaussian loglikelihood, the distribution of residuals normalized by the true, input errors for the true underlying model evaluated at the true parameters is Gaussian with zero mean and unity variance; i.e. the true model removes all foreground power, leaving exactly only Gaussian noise at the input level. All other models will produce a distribution of residuals which will deviate, whether slightly or significantly, from the case where the errors are known and the model is the truth.

One can quantify this deviation by comparing the cumulative distribution function (CDF) of a particular model fit's normalized residuals $\delta / \sigma_n$ to the true model's normalized residuals, which is the CDF of a unit normal distribution, using the KS-test. The latter produces a p-value for a particular model fit, $p_{ks}$. If the null hypothesis is that a particular model produces normalized residuals which follow a unit normal distribution (i.e. the \textit{true} distribution), then we reject the null hypothesis if the particular model's fit produces a p-value below some threshold. The model does not follow a unit normal distribution in that case, and we conclude that the foreground has not been removed to the noise level\footnote{One could also employ the Anderson-Darling test in a similar manner as the KS-test, as it will be more sensitive to deviations other than those in the median. However, the AD statistic for a given significance level only specifies a threshold for when the null hypothesis may be rejected, and does not necessarily delineate model preference between two models which are both beneath this threshold. We leave such testing for future work.}. To be clear, here we interpret the $p_{ks}$-value from the KS-test to mean the probability of observing the KS-test statistic $D$ given that the null hypothesis (the model residuals follow a unit normal distribution) is true. If the $p_{ks}$ value falls below some threshold of significance, then we conclude that there is a low probability that the null hypothesis is true, or equivalently, a low probability that the model residuals follow a unit normal distribution. We choose our default threshold to be $p_{ks} < 0.05$; we leave a full quantification of the best threshold for determining goodness-of-fit to future work, although see Section
\ref{sec:p-values} for a brief discussion on this value.

Therefore, models which produce $p_{ks} > 0.05$ pass the null hypothesis, and their residuals are indistinguishable from the true model's residuals. Furthermore, if model $A$ produces a p-value of $p_{a}$ and model B produces a p-value of $p_{b}$, both pass the null hypothesis test, and $p_a > p_b$, we conclude that $p_a$ is a better fit to the data than $p_b$.

\subsection{Testing Model Inputs}
\subsubsection{Ideal Model Inputs}
The nonlinear and linear foreground models rely on explicit spatial forms for the temperature BM, the spectral index PM, and the beam. When these model inputs are identical to those used to compute the REACH mock, then model fits test to what noise level the REACH mock can be fit in the ideal case. In this case we write $T_0 = T_{408}$ and $\boldsymbol{M}_j(\beta_R)$, as the BM and the spectral index masks are taken directly from those used to generate the intrinsic foreground, $T_{408}$ and $\beta_R$, respectively.

Care still has to be taken, however, to choose the correct number of regions $N_r$ and nonlinear parameters for the optimal fit. In general, the number of regions which maximizes the Bayesian evidence is a function of both the size of the beam and the number of LST bins in $T_n$. The former enters into consideration because larger beams will average more of the sky together and make many small regions largely indistinguishable from one large region, while the latter also enforces spatial resolution through time-dependence. More LST bins requires greater spatial resolution, although this too is limited by the size of the beam. See Appendix \ref{app-skyregions} for a demonstration of this effect in the case of the linear-forward models. 

Therefore, we appeal to the Bayesian evidence to choose the optimal number of regions $N_r$ for both the nonlinear and linear models. We find that for the nonlinear models the number of regions which maximizes the Bayesian evidence is as follows: $N_r = 4$ for 1 LST bin, $N_r = 8$ for 2 LST bins, $N_r = 9$ for 5 LST bins, and $N_r = 9$ also for 10 LST bins, in the ideal model input case. Due to the computational expense of running each of these models for many regions, we use these optimal numbers for all of our fits.

\begin{figure*}
    \centering
    \includegraphics[width=0.96\textwidth]{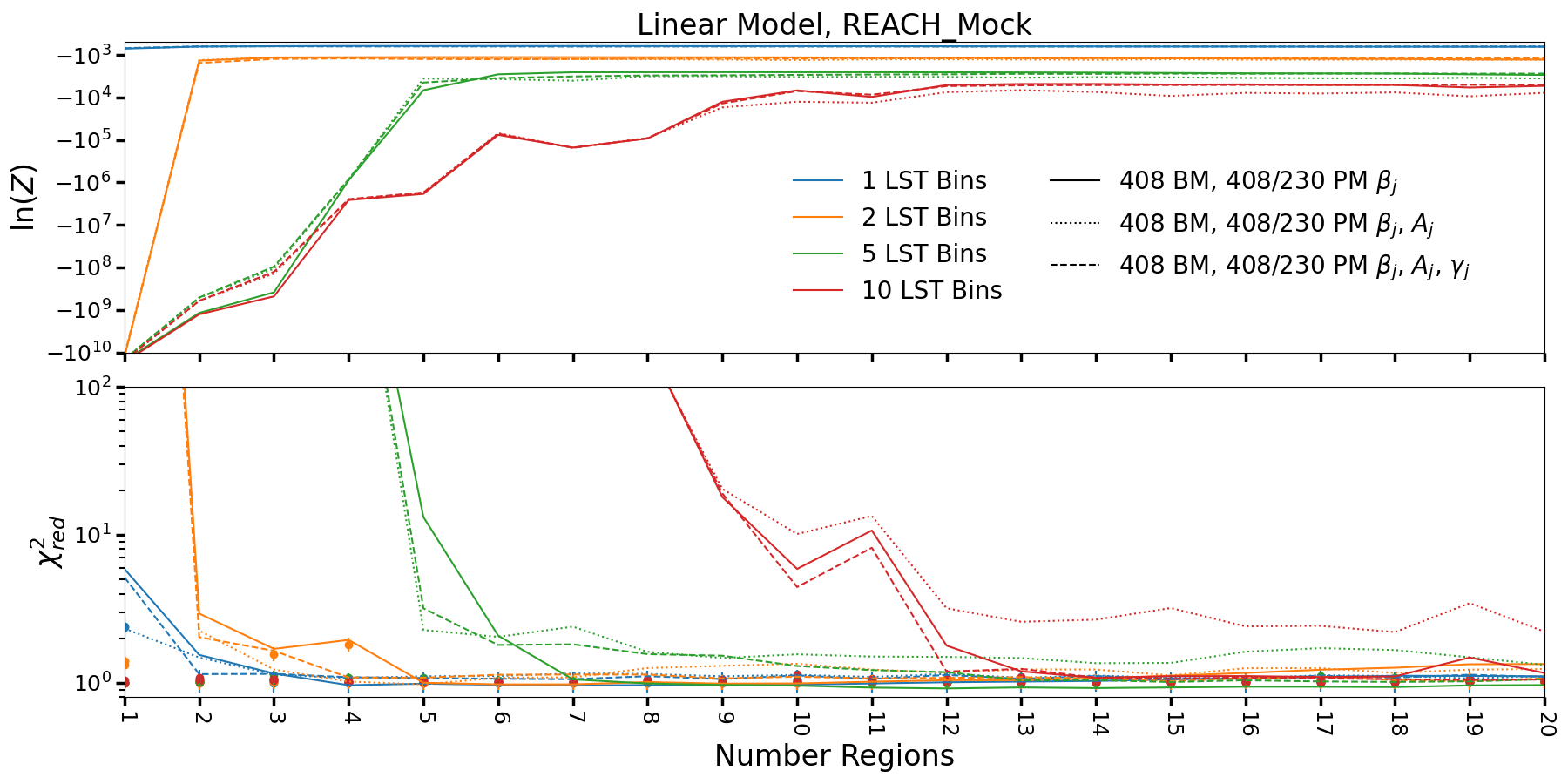}
    \caption{The linear model evidences $\ln{Z}$ and $\chi^2_{red}$ for fits to different numbers of LST bins and regions $N_r$. The models shown here (solid, dashed, and dotted) are when the model inputs (BM, PM) are the same as those in the REACH mock, the ideal case, but the number of nonlinear parameters included varies. Each LST bin tends to saturate at a maximum number of regions, with greater LST bin mock foregrounds requiring greater spatial resolution. The solid, colored dots in the bottom panel correspond to the expectation value $E[\chi^2_{red}]$ for each model input and number of LST bins. Similar analyses are carried out when the model inputs differ from the REACH mock.}
    \label{fig:lin-model-evidence-chi}
\end{figure*}

For the linear model, we compute the Bayesian evidence for $N_r = 1 - 20$ for each number of LST bins and choose the model with the highest evidence. If two models exhibit a similar evidence $\ln{Z_1} - \ln{Z_2} \leq 3$, following \cite{kass_bayes_1995}, we select the model input with $\chi^2_{red}$ which is the least number of $\sigma$ away from the expected reduced chi-squared, $E[\chi^2_{red}]$. If several model inputs have similar evidences and number of sigma, we select the model input that has the smallest $N_r$. 

We also use this same method of model parameter number selection for the polynomial and maximally smooth models.

Figure \ref{fig:lin-model-evidence-chi} shows both the evidence $\ln{Z}$ and $\chi^2_{red}$ for each number of LST bins (blue, orange, green, and red for 1, 2, 5, and 10, respectively), and when different nonlinear model parameters are included in the matrix of eigenspectra $\boldsymbol{F}_{\text{fg}}$ (solid, dotted, and dashed lines for $[\beta_j]$, $[\beta_j, A_j]$, $[\beta_j, A_j, \gamma_j]$, respectively). It is interesting to note that the evidence for the linear models saturates at a given number of regions, dependent upon the number of LST bins. In general, one region is typically enough for 1 LST bin, while at least 2 regions are required for 2 LST bins, 5 regions for 5 LST bins, and 12 regions for 10 LST bins. We have checked that this pattern is consistent across different model inputs (BM, PM) and numbers of nonlinear parameters included. This trend reflects the fact mentioned above that more LST bins requires higher spatial resolution (see Appendix \ref{app-skyregions}). The filled circle symbols in the bottom panel of Figure \ref{fig:lin-model-evidence-chi} mark the expected values $E[\chi^2_{red}]$ for each fit.

\subsubsection{Incorrect Model Inputs}
To test the dependence of these models on their inputs, in addition to the ideal case (mathematically represented by $\boldsymbol{M}_j(\beta_R) = \boldsymbol{M}_j(\beta_{408/230})$ and $T_{0} = T_{408}$), we compute nonlinear and linear fits when one of the inputs $T_0$ or $\boldsymbol{M}_j$ is incorrect, and leave tests of the beam for future work. We wish to use model inputs which are similar to those used in the REACH mock, but different enough to test the limits of the models. For the case when the BM is incorrect, we set $T_0 = T_{300}$; for the case when the PM is incorrect, we set $\boldsymbol{M}_j = \boldsymbol{M}_j(\beta_{300/130})$, where the spectral index map $\beta_{300/130}$ is the ratio of the GSM maps generated at 300 and 130 MHz.

\begin{figure*}
    \centering
        \includegraphics[width=0.49\textwidth]{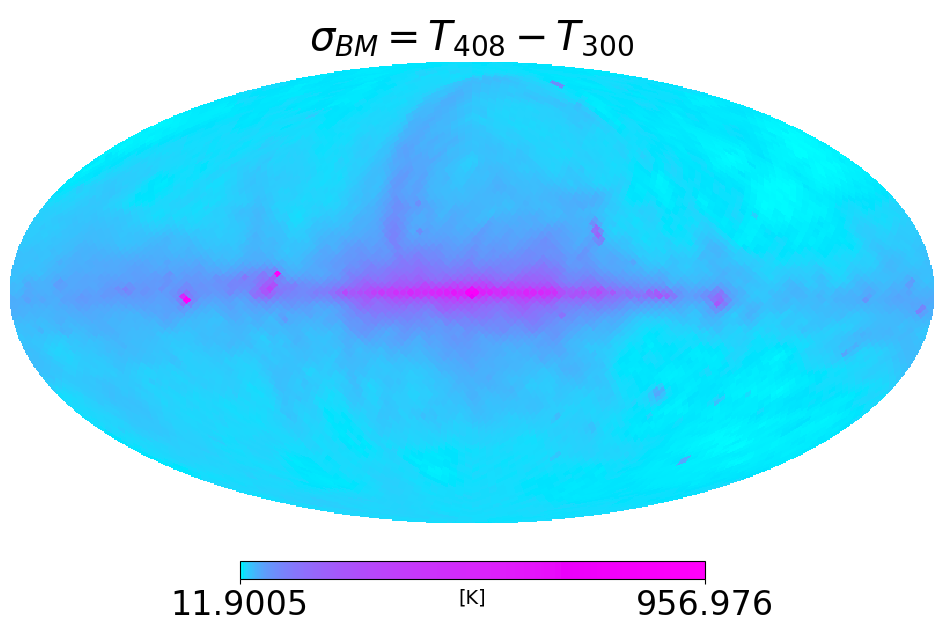}
    \includegraphics[width=0.49\textwidth]{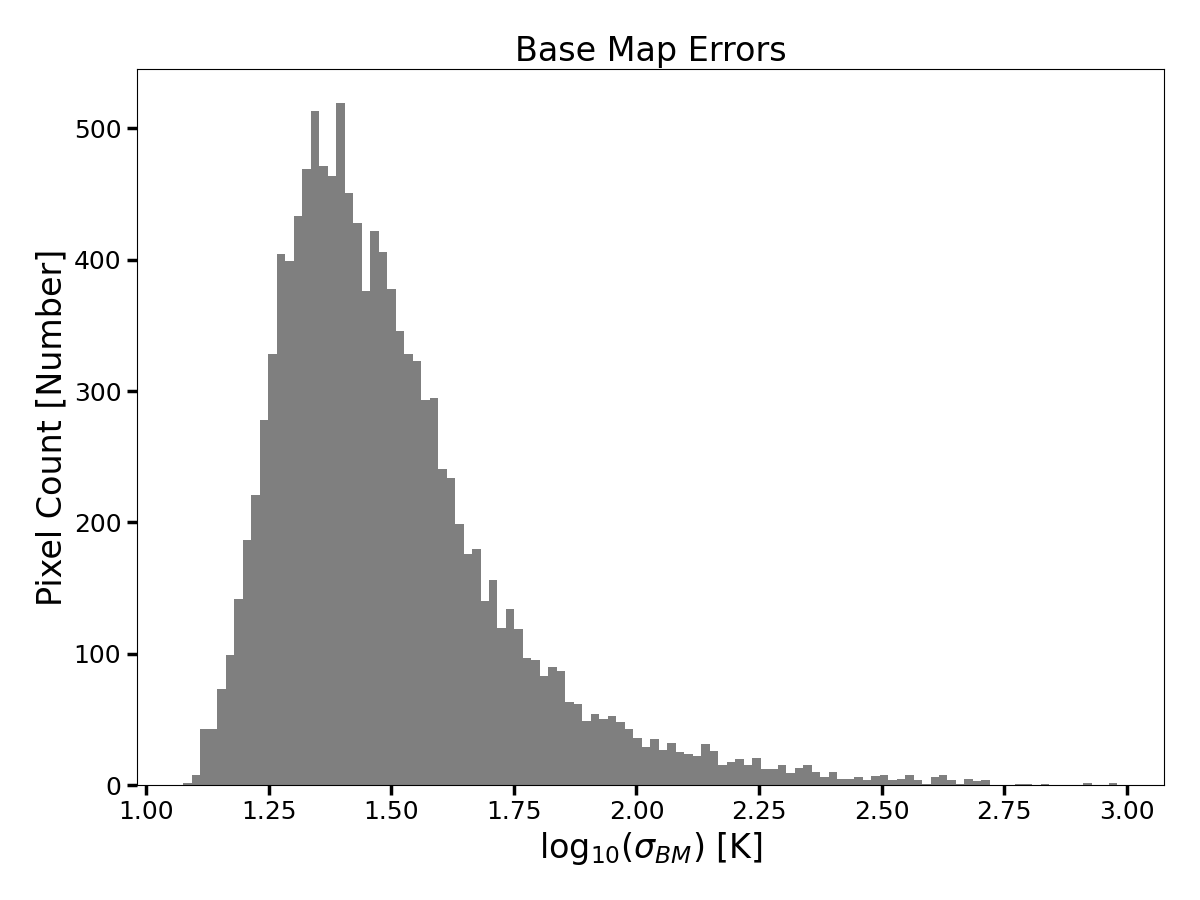}
    \caption{Mollvweide view and histogram of the difference between the temperature BM used to create the REACH mock ($T_{408}$) and the model input ($T_{300}$). The right panel shows the temperature difference per pixel, which peaks around 12-17 K, while the left panel shows the view on the galactic-centered sky. Pixels which are cyan have model input values very close to the REACH mock, while those in magenta are more different. Note that these errors still trace out the features of the galaxy, and are thus indicative of errors which are proportional to the temperatures in each pixel. Hence, the pixel errors near the galactic center, which is brighter, are larger.}
    \label{fig:bm-test}
\end{figure*}

The left panel of Figure \ref{fig:bm-test} shows a mollweide view of the absolute difference between the $T_{408}$ BM of the REACH mock and the $T_{300}$ BM model input. The right panel of the figure shows a histogram of the pixel temperatures, which can be thought of as the errors between the model input temperature map and the temperature map in the mock. Note that these represent differences in spatial structure in the assumed BM. These errors, which trace the galactic structure, are analogous to errors in the temperature map which are proportional to the temperature in each pixel and with zero degrees of correlation across the sky. See also \cite{pagano_general_2022} for similar error map analyses. 

For clarity, note that these figures show the absolute temperature difference between the two maps: if the pixels (and thus spatial structure) of one map were identical up to a multiplicative factor, then we would expect no difference when changing the model input map, as the spatial structures would be identical and the scaling by the reference frequency $\nu_0$ would account for the absolute temperature difference. This test examines how small pixel-by-pixel differences in the BM affect the ability of the foreground model to account for them, generally through using more parameters (like the magnitude parameters, $A_j$). To summarize, we are asking the question, if the true pixel absolute values are those given by $T_{408}$, but we use a model input with slightly different spatial structure in the temperature map, such as $T_{300}$, how many parameters will it take for the foreground to fit the mock spectra using the incorrect model input down to the noise level?

\begin{figure*}
    \centering
    \includegraphics[width=0.48\textwidth]{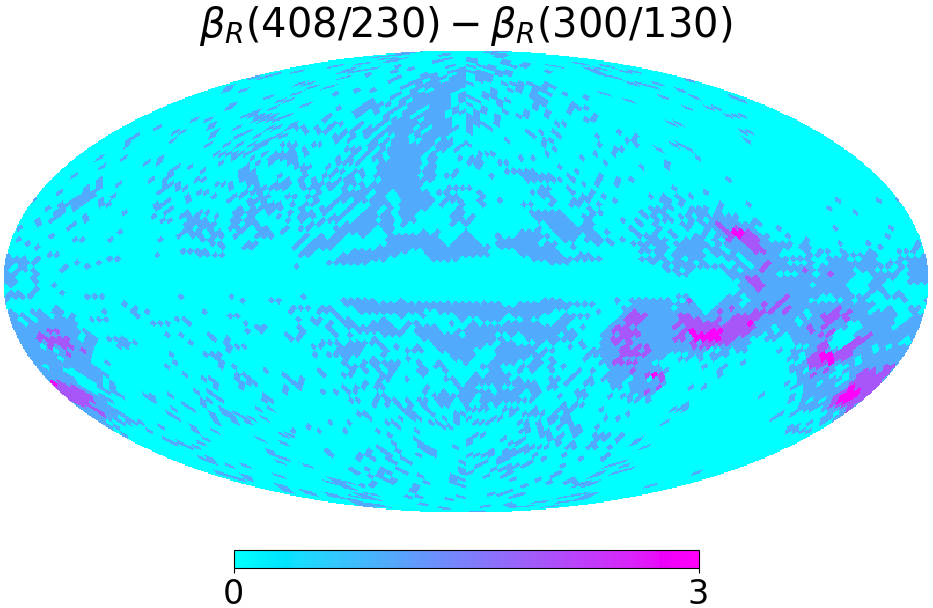}
    \includegraphics[width=0.48\textwidth]{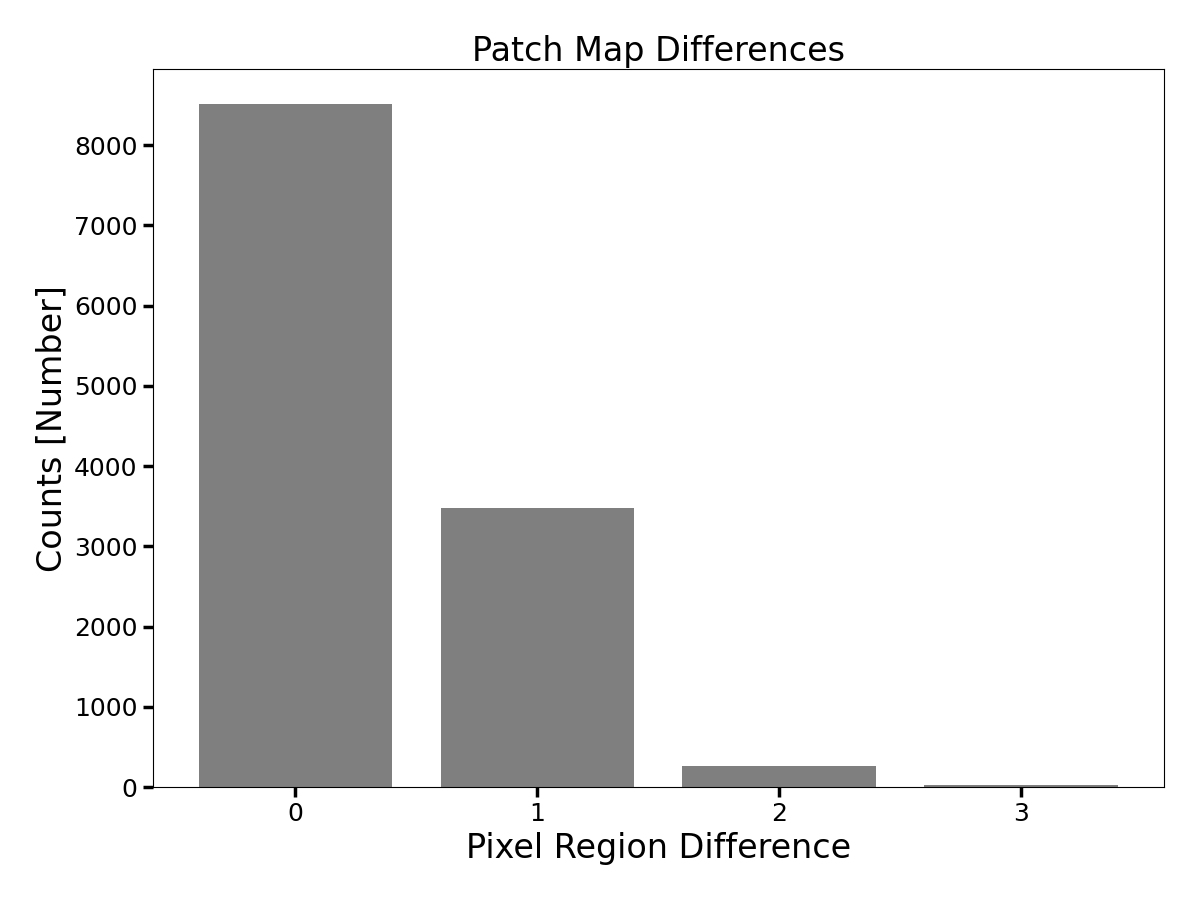}
    \caption{Mollweide view (left panel) and histogram (right panel) of the difference between the spectral index patch map used to create the REACH mock (408/230) and the patch map used as a model input (300/130). Each region in ascending order is assigned a number between 0 and 7, before the difference is taken. Pixels which are in the same region in both the mock and model input are shown in cyan in the left panel and have a difference value of zero, while those which differ by one region have a value of one, and so on.}
    \label{fig:pm-test}
\end{figure*}

Similarly, Figure \ref{fig:pm-test} shows an example mollweide view of the difference in the two patch maps for $N_r = 8$ when each region is assigned a number from zero to seven. Therefore, a difference of zero (cyan color) means that both pixels are assigned to the same patch of the sky, while a difference of one means that they differ by one region, and so on. The right panel of Figure \ref{fig:pm-test} shows a bar chart of the pixels, where one can see that roughly seventy percent of the pixels are within the same region, and twenty-eight percent differ by a single region on the sky. We do not consider this too draconian a difference in patch assignations, as seen by the preponderance of cyan in the left panel. 

All together then, for the nonlinear and linear models, we perform three categories of tests: when the model inputs (BM, PM) are ideal, when the BM only is incorrect, and finally when the PM only is incorrect. For the polynomial models, we only test to what residual level they can fit the single LST bin REACH mock for various numbers of polynomial parameters $N_{py}$ or $N_{MSF}$.

\section{Results}
\label{sec-results}

\begin{table*}[t]
    \setlength{\tabcolsep}{0.27em}
    \def\arraystretch{0.95}
    \centering
    \begin{tabular}{c|cc|ccccccc|ccccc} 
        \hline
        & \multicolumn{2}{c}{Model} & \multicolumn{7}{|c|}{Linear} & \multicolumn{5}{c}{Nonlinear} \\
        \hline
        \hline
        LSTs & Input & $\theta_j$ & $N_r$ & $\chi^2_{red}$ & $\sigma$ & $\ln{Z}$ & $N_x$ & $p_{ks}$ & BF & $N_r$ & $\ln{Z}$ & $N_{\theta}$ & $p_{ks}$ & BF \\
        \hline
        \hline
        \multirow{9}{*}{1} & \multirow{3}{*}{IDEAL} & $\beta_j$ & 4 & 0.96 & 1 & -615.71 & 7 & 0.94 & -- & \cellcolor{Goldenrod}4 & 
        \cellcolor{Goldenrod}{-81.05} & \cellcolor{Goldenrod}{4} & \cellcolor{Goldenrod}{0.99} & \cellcolor{Goldenrod}{NL1} \\
        & & $\beta_j$, $A_j$ & 3 & 1.14 & 1 & -625.24 & 6 & 0.97 & -- & 4 &
        -105.78 & 8 & 0.97 & -- \\
        & & $\beta_j$, $A_j$, $\gamma_j$ & 2 & 1.14 & 1 & -628.66 & 6 & 0.90 & -- & 4 & 
        -119.07 & 12 & 0.88 & -- \\
        \cline{2-15}
        & \multirow{3}{*}{300 BM} & $\beta_j$ & 4 & 0.96 & 1 & -614.26 & 7 & 0.93 & -- & \cellcolor{lightgray} 4 & 
        \cellcolor{lightgray} -7.8e4 & \cellcolor{lightgray} 4 & \cellcolor{lightgray} 1.36e-24 & \cellcolor{lightgray} -- \\ 
        & & $\beta_j$, $A_j$ & \cellcolor{Goldenrod}{3} & \cellcolor{Goldenrod}{1.10} & \cellcolor{Goldenrod}{1} & \cellcolor{Goldenrod}{-623.23} & \cellcolor{Goldenrod}{6} & \cellcolor{Goldenrod}{0.99} & \cellcolor{Goldenrod}{L1} & 4 & 
        -107.39 & 8 & 0.98 & -- \\
        & & $\beta_j$, $A_j$, $\gamma_j$ & 2 & 1.12 & 1 & -626.91 & 6 & 0.95 & -- & 4 &
        -98.22 & 12 & 0.93 & -- \\
        \cline{2-15}
        & \multirow{3}{*}{(300/130) PM} & $\beta_j$ & 4 & 0.99 & 1 & -613.8 & 5 & 0.998 & -- & \cellcolor{lightgray} 4 & 
        \cellcolor{lightgray} -1.74e12 & \cellcolor{lightgray} 4 & \cellcolor{lightgray} $\sim 0$ & \cellcolor{lightgray} -- \\
        & & $\beta_j$, $A_j$ & 3 & 1.06 & 1 & -618.83 & 8 & 0.998 & -- & 4 &
        -112.26 & 8 & 0.94 & -- \\
        & & $\beta_j$, $A_j$, $\gamma_j$ & 2 & 1.09 &
        1 & -623.53 & 6 & 0.88 & -- & \cellcolor{Goldenrod}{4} & 
        \cellcolor{Goldenrod}{-112.71} & \cellcolor{Goldenrod}{12} & \cellcolor{Goldenrod}{0.999} & \cellcolor{Goldenrod}{NL2} \\
        \hline
        \hline
        \multirow{9}{*}{2} & \multirow{3}{*}{IDEAL} & $\beta_j$ & 5 & 1.00 & 1 & -1129.88 & 9 & 0.68 & -- & 8 &
        -154.24 & 8 & 0.81 & -- \\
        & & $\beta_j$, $A_j$ & 4 & 1.02 & 1 & -1160.86 & 13 & 0.81 & -- & \cellcolor{Goldenrod}{8} & 
        \cellcolor{Goldenrod}{-335.72} & \cellcolor{Goldenrod}{16} & \cellcolor{Goldenrod}{0.83} & \cellcolor{Goldenrod}{NL3}\\
        & & $\beta_j$, $A_j$, $\gamma_j$ & 15 & 1.03 & 1 & -1192.70 & 18 & 0.80 & -- & 8 & 
        -307.33 & 24 & 0.09 & -- \\
        \cline{2-15}
        & \multirow{3}{*}{300 BM} & $\beta_j$ & 5 & 1.03 & 1 & -1130.74 & 13 & 0.51 & -- & 8 \cellcolor{lightgray} & 
        \cellcolor{lightgray} -6.96e6 & \cellcolor{lightgray} 8 & \cellcolor{lightgray} 5.6e-45 & \cellcolor{lightgray} -- \\ 
        & & $\beta_j$, $A_j$ & \cellcolor{Goldenrod}{4} & \cellcolor{Goldenrod}{1.02} & \cellcolor{Goldenrod}{1} & \cellcolor{Goldenrod}{-1155.56} & \cellcolor{Goldenrod}{13} & \cellcolor{Goldenrod}{0.89} & \cellcolor{Goldenrod}{L2} & 8 & 
        -364.17 & 16 & 0.82 & -- \\
        & & $\beta_j$, $A_j$, $\gamma_j$ & 4 & 1.08 & 1 & -1189.71 & 14 & 0.78 & -- & 8 & 
        -234.77 & 24 & 0.50 & -- \\
        \cline{2-15}
        & \multirow{3}{*}{(300/130) PM} & $\beta_j$ & 6 & 1.13 & 1 & -1134.67 & 21 & 0.71 & -- & \cellcolor{lightgray} 8 & 
        \cellcolor{lightgray} -1.82e12 & \cellcolor{lightgray} 8 & \cellcolor{lightgray} $\sim 0$ & \cellcolor{lightgray} -- \\
        & & $\beta_j$, $A_j$ & \cellcolor{Goldenrod}{4} & \cellcolor{Goldenrod}{1.00} & \cellcolor{Goldenrod}{1} & \cellcolor{Goldenrod}{-1142.86} & \cellcolor{Goldenrod}{13} & \cellcolor{Goldenrod}{0.97} & \cellcolor{Goldenrod}{L3} & \cellcolor{lightgray} 8 & 
        \cellcolor{lightgray} -500.73 & \cellcolor{lightgray} 16 & \cellcolor{lightgray} 0.002 & \cellcolor{lightgray} -- \\
        & & $\beta_j$, $A_j$, $\gamma_j$ & 4 & 1.04 & 1 & -1162.65 & 14 & 0.96 & -- & 8 & 
        -220.24 & 24 & 0.26 & -- \\
        \hline
        \hline
        \multirow{9}{*}{5} & \multirow{3}{*}{IDEAL} & $\beta_j$ & 8 & 0.99 & 1 & -2553.1 & 25 & 0.85 & -- & \cellcolor{lightgray} 9 & 
        \cellcolor{lightgray} -2034.15 & \cellcolor{lightgray} 9 & \cellcolor{lightgray} 1.1e-23 & \cellcolor{lightgray} --\\
        & & $\beta_j$, $A_j$ & 9 & 1.47 & 7 & -3178.06 & 30 & 0.14 & -- & 9 &
        -391.24 & 18 & 0.87 & -- \\
        & & $\beta_j$, $A_j$, $\gamma_j$ & 18 & 1.01 & 1 & -2746.97 & 37 & 0.85 & -- & \cellcolor{Goldenrod}{9} & 
        \cellcolor{Goldenrod}{-506.06} & \cellcolor{Goldenrod}{27} & \cellcolor{Goldenrod}{0.96} & \cellcolor{Goldenrod}{NL4} \\
        \cline{2-15}
        & \multirow{3}{*}{300 BM} & $\beta_j$ & \cellcolor{Goldenrod}{8} & \cellcolor{Goldenrod}{1.07} & \cellcolor{Goldenrod}{1} & \cellcolor{Goldenrod}{-2570.87} & \cellcolor{Goldenrod}{24} & \cellcolor{Goldenrod}{0.97} & \cellcolor{Goldenrod}{L4} & \cellcolor{lightgray} 9 & 
        \cellcolor{lightgray} -3.27e7 & \cellcolor{lightgray} 9 & \cellcolor{lightgray} 2.5e-127 & \cellcolor{lightgray} -- \\ 
        & & $\beta_j$, $A_j$ & 9 & 1.45 & 7 & -3110.48 & 30 & 0.19 & -- & 9 & 
        -467.83 & 18 & 0.58 & -- \\
        & & $\beta_j$, $A_j$, $\gamma_j$ & 18 & 1.01 & 1 & -2789.27 & 35 & 0.73 & -- & 9 & 
        -657.18 & 27 & 0.19 & -- \\
        \cline{2-15}
        & \multirow{3}{*}{(300/130) PM} & $\beta_j$ & 14 & 1.03 & 1 & -2642.16 & 30 & 0.86 & -- & \cellcolor{lightgray} 9 & 
        \cellcolor{lightgray} -1.83e12 & \cellcolor{lightgray} 9 & \cellcolor{lightgray} $\sim 0$ & \cellcolor{lightgray} -- \\
        & & $\beta_j$, $A_j$ & 10 & 1.31 & 5 & -2773.33 & 30 & 0.68 & -- & \cellcolor{lightgray} 9 & 
        \cellcolor{lightgray} -2.04e5 & \cellcolor{lightgray} 18 & \cellcolor{lightgray} 1.2e-96 & \cellcolor{lightgray} -- \\
        & & $\beta_j$, $A_j$, $\gamma_j$ & \cellcolor{Goldenrod}{12} & \cellcolor{Goldenrod}{1.01} & \cellcolor{Goldenrod}{1} & \cellcolor{Goldenrod}{-2662.62} & \cellcolor{Goldenrod}{33} & \cellcolor{Goldenrod}{0.93} & \cellcolor{Goldenrod}{L5} & \cellcolor{lightgray} 9 & 
        \cellcolor{lightgray} -2221.81 & \cellcolor{lightgray} 27 & \cellcolor{lightgray} 8.6e-20 & \cellcolor{lightgray} -- \\
        \hline
        \hline
        \multirow{9}{*}{10} & \multirow{3}{*}{IDEAL} & $\beta_j$ & 14 & 1.08 & 2 & -4833.53 & 36 & 0.80 & -- & 9 \cellcolor{lightgray} & 
        \cellcolor{lightgray} -2.20e5 & \cellcolor{lightgray} 9 & \cellcolor{lightgray} 1.4e-198 & \cellcolor{lightgray} -- \\
        & & $\beta_j$, $A_j$ & \cellcolor{lightgray} 13 & \cellcolor{lightgray} 2.58 & \cellcolor{lightgray} 33 & \cellcolor{lightgray} -6783.17 & \cellcolor{lightgray} 46 & \cellcolor{lightgray} 1.91e-8 & \cellcolor{lightgray} -- & \cellcolor{lightgray} 9 & 
        \cellcolor{lightgray} -2.9e4 & \cellcolor{lightgray} 18 & \cellcolor{lightgray} 7.1e-148 & \cellcolor{lightgray} -- \\
        & & $\beta_j$, $A_j$, $\gamma_j$ & \cellcolor{Goldenrod}{19} & \cellcolor{Goldenrod}{1.04} & \cellcolor{Goldenrod}{1} & \cellcolor{Goldenrod}{-5059.1} & \cellcolor{Goldenrod}{58} & \cellcolor{Goldenrod}{0.82} & \cellcolor{Goldenrod}{L6} & \cellcolor{lightgray} 9 & 
        \cellcolor{lightgray} -1.29e4 & \cellcolor{lightgray} 27 & \cellcolor{lightgray} 2.5e-107 & \cellcolor{lightgray} -- \\
        \cline{2-15}
        & \multirow{3}{*}{300 BM} & $\beta_j$ & \cellcolor{lightgray} 12 & \cellcolor{lightgray} 1.40 & \cellcolor{lightgray} 8 & \cellcolor{lightgray} -5004.78 & \cellcolor{lightgray} 39 & \cellcolor{lightgray} 0.03 & \cellcolor{lightgray} -- & \cellcolor{lightgray} 9 & 
        \cellcolor{lightgray} -3.54e7 & \cellcolor{lightgray} 9 & \cellcolor{lightgray} 2.4e-252 & \cellcolor{lightgray} -- \\ 
        & & $\beta_j$, $A_j$ & \cellcolor{lightgray} 13 & \cellcolor{lightgray} 2.39 & \cellcolor{lightgray} 29 & \cellcolor{lightgray} -6696.14 & \cellcolor{lightgray} 46 & \cellcolor{lightgray} 1.3e-7 & \cellcolor{lightgray} -- & \cellcolor{lightgray} 9 & 
        \cellcolor{lightgray} -1.48e4 & \cellcolor{lightgray} 18 & \cellcolor{lightgray} 1.7e-122 & \cellcolor{lightgray} -- \\
        & & $\beta_j$, $A_j$, $\gamma_j$ & \cellcolor{Goldenrod}{19} & \cellcolor{Goldenrod}{1.07} & \cellcolor{Goldenrod}{2} & \cellcolor{Goldenrod}{-5183.84} & \cellcolor{Goldenrod}{56} & \cellcolor{Goldenrod}{0.94} & \cellcolor{Goldenrod}{L7} & \cellcolor{lightgray} 9 & 
        \cellcolor{lightgray} -9134.45 & \cellcolor{lightgray} 27 & \cellcolor{lightgray} 8.6e-80 & \cellcolor{lightgray} -- \\
        \cline{2-15}
        & \multirow{3}{*}{(300/130) PM} & $\beta_j$ & \cellcolor{lightgray} 17 & \cellcolor{lightgray} 2.34 & \cellcolor{lightgray} 28 & \cellcolor{lightgray} -6408.78 & \cellcolor{lightgray} 54 & \cellcolor{lightgray} 2.7e-9 & \cellcolor{lightgray} -- & \cellcolor{lightgray} 9 & 
        \cellcolor{lightgray} -1.84e12 & \cellcolor{lightgray} 9 & \cellcolor{lightgray} $\sim 0$ & \cellcolor{lightgray} -- \\
        & & $\beta_j$, $A_j$ & \cellcolor{lightgray} 20 & \cellcolor{lightgray} 2.02 & \cellcolor{lightgray} 5 & \cellcolor{lightgray} -6060.94 & \cellcolor{lightgray} 73 & \cellcolor{lightgray} 1.0e-4 & \cellcolor{lightgray} -- & \cellcolor{lightgray} 9 & 
        \cellcolor{lightgray} -1.90e7 & \cellcolor{lightgray} 18 & \cellcolor{lightgray} 2.0e-251 & \cellcolor{lightgray} -- \\
        & & $\beta_j$, $A_j$, $\gamma_j$ & \cellcolor{Goldenrod}{20} & \cellcolor{Goldenrod}{1.34} & \cellcolor{Goldenrod}{7} & \cellcolor{Goldenrod}{-5104.81} & \cellcolor{Goldenrod}{79} & \cellcolor{Goldenrod}{0.12} & \cellcolor{Goldenrod}{L8} & \cellcolor{lightgray} 9 & 
        \cellcolor{lightgray} -8.18e6 & \cellcolor{lightgray} 27 & \cellcolor{lightgray} 6.6e-223 & \cellcolor{lightgray} -- \\
        \hline
    \end{tabular}
    \caption{Linear and Nonlinear model fit results to the REACH mock. Each category of fits is delineated by its model inputs (IDEAL, BM, or PM) and separated by a dividing horizontal line, within a given number of LSTs. IDEAL means the model input is the same as the REACH mock, BM denotes the temperature base map used in the model input which is different from the REACH Mock, and PM denotes the spectral index patch used in the model input which differs from the REACH mock. Within a category, the best-fitting model based upon the KS-test p-value $p_{ks}$ are outlined in gold. Model fits which do not pass the null hypothesis exhibit $p_{ks}<0.05$ and are outlined in gray. The last column $BF$ indicates the best-fitting model within each category, labelled according to whether it is linear $L$ or nonlinear $NL$.}
    \label{tab:lin-and-nonlin-results}
\end{table*}

\begin{table}[t]
    \setlength{\tabcolsep}{0.27em}
    \def\arraystretch{0.95}
    \centering
    \begin{tabular}{c|cccccc}
        \hline
        \hline
         Model & $N_{py}$ & $\chi^2_{red}$ & $\sigma$ & $\ln{Z}$ & RMS (mK) & $p_{ks}$ \\
         \hline
         \hline
         \multirow{3}{*}{LinLogPoly} & \cellcolor{lightgray} 5 & \cellcolor{lightgray} 6.2 & \cellcolor{lightgray} 34 & \cellcolor{lightgray} -310 & \cellcolor{lightgray} 31 & \cellcolor{lightgray} 9.3e-6 \\
         & 6 & 1.44 & 3 & -104 & 13 & 0.40 \\
         & 7 & 1.38 & 2.5 & -103 & 12 & 0.36 \\
         & 8 & 1.38 & 2.5 & -108 & 12 & 0.4 \\
         & \cellcolor{Goldenrod} 9 & \cellcolor{Goldenrod} 1.33 & \cellcolor{Goldenrod} 2.1 & \cellcolor{Goldenrod} -102 & \cellcolor{Goldenrod} 12 & \cellcolor{Goldenrod} 0.67 \\
         & 10 & 1.21 & 1.39 & -115 & 11 & 0.63 \\
         \hline
         \multirow{3}{*}{LinPoly} & \cellcolor{lightgray} 5 & \cellcolor{lightgray} 2.85 & \cellcolor{lightgray} 12 & \cellcolor{lightgray} -163 & \cellcolor{lightgray} 16 & \cellcolor{lightgray} 0.04 \\
         & 6 & 1.60 & 4 & -111 & 14 & 0.33 \\
         & 7 & 1.43 & 2.8 & -109 & 13 & 0.61 \\
         & 8 & 1.43 & 2.8 & -108 & 12 & 0.39 \\
         & 9 & 1.28 & 1.8 & -104 & 12 & 0.71 \\
         & \cellcolor{Goldenrod} 10 & \cellcolor{Goldenrod} 1.28 & \cellcolor{Goldenrod} 1.8 & \cellcolor{Goldenrod} -121 & \cellcolor{Goldenrod} 12 & \cellcolor{Goldenrod} 0.77 \\
         \hline
         LinPhys & \cellcolor{lightgray} 5 & \cellcolor{lightgray} 9.04 & \cellcolor{lightgray} 53 & \cellcolor{lightgray} -4.3e7 & \cellcolor{lightgray} 43 & \cellcolor{lightgray} 4.08e-29 \\
         \hline
    \end{tabular}
    \caption{Polynomial Model fits to the REACH mock. The color codes here are the same as for the nonlinear and linear model fits in Table~\ref{tab:lin-and-nonlin-results}.}
    \label{tab:poly-fits}
\end{table}

\begin{table}[ht]
    \setlength{\tabcolsep}{0.25em}
    \def\arraystretch{0.9}
    \centering
    \begin{tabular}{c|cccc}
        \hline
        \hline
         Model & $N_{MSF}$ & RMS (mK) & $p_{ks}$ \\
         \hline
         \hline
         \multirow{4}{*}{Diff Poly} & \cellcolor{lightgray} 5 & 
         \cellcolor{lightgray} 6e3 & \cellcolor{lightgray} 1.4e-24 \\
         & \cellcolor{lightgray} 6 & 
         \cellcolor{lightgray} 1.36e3 & \cellcolor{lightgray} 1.56e-21 \\
         & \cellcolor{Goldenrod} 10 & 
         \cellcolor{Goldenrod} 12 & \cellcolor{Goldenrod} 0.76 \\
         & 15 &
         11 & 0.58 \\
         \hline
         \multirow{3}{*}{LogLog Poly} & \cellcolor{lightgray} 5 & 
         \cellcolor{lightgray} 2.1e4 & \cellcolor{lightgray} 8.9e-27 \\
         & \cellcolor{lightgray} 6 & 
         \cellcolor{lightgray} 2.2e3 & \cellcolor{lightgray} 8.9e-27 \\
         & \cellcolor{lightgray} 10 & 
         \cellcolor{lightgray} 103 & \cellcolor{lightgray} 1.4e-17 \\
         & \cellcolor{lightgray} 15 & 
         \cellcolor{lightgray} 23 & \cellcolor{lightgray} 0.008 \\
         \hline
    \end{tabular}
    \caption{Maximally Smooth Polynomial fits to the REACH mock. The color codes here are the same as for the nonlinear and linear model fits in Table~\ref{tab:lin-and-nonlin-results}.}
    \label{tab:msf-fits}
\end{table}

Tables \ref{tab:lin-and-nonlin-results} \ref{tab:poly-fits}, and \ref{tab:msf-fits} summarize the main models, evidences, and goodness-of-fit statistics (such as $\chi^2_{red}$ and the KS test p-value, $p_{ks}$) produced when fitting the REACH mock with linear and nonlinear, polynomial, and maximally smooth foreground models, respectively. Each model is defined first by its model inputs (such as LSTs, Input Base Map, Patch Map, and parameters), and secondly by the number of parameters in the fit, $N_{\Theta}$, where $\Theta$ takes on the symbols $[x, \theta, py, MSF]$ for linear, nonlinear, polynomial, and maximally smooth models, respectively. Where the model is linear in its parameters, such as for the linear and polynomial foreground models, we also include the number of sigma $\sigma$ the model fit lies away from the expected $\chi^2_{red}$, for the linear models where the degrees of freedom are calculable. A value of $\sigma = 1$, therefore, means that the value of $\chi^2_{red}$ produced by the model fit lies within one sigma of $E[\chi^2_{red}]$.

We color-code each table, for a particular model or model input (as delineated by the horizontal lines), according to its $p_{ks}-$value. Under the null hypothesis, the noise-weighted residuals produced by a particular model fit do not deviate significantly from a Gaussian-distribution with mean $\mu = 0$ and variance $\sigma^2 = 1$. Therefore, if a fit produces a $p_{ks} < 0.05$, we reject the null hypothesis and conclude that residuals deviate from a unit normal Gaussian; i.e. that significant foreground power remains in the residuals at or above the noise level in the data. Models with the highest $p_{ks}$-value (and therefore the best-fit) for a particular input are colored in \colorbox{Goldenrod}{gold}, while models which fail the null hypothesis are colored in \colorbox{lightgray}{gray}. Additionally, for the linear and nonlinear foreground models, we label the best-fitting gold model with either $L$ or $NL$, and a number, as seen in the column labeled $BF$ for Best-Fit.

\subsection{Nonlinear and Linear Models}
As was mentioned previously, neither $\chi^2_{red}$ nor the $\ln{Z}$ are always jointly representative of the best-fit model for the linear and nonlinear foreground models. The latter allows one to form a Bayes ratio and hence calculate model preference, while the former describes goodness-of-fit when the model is linear. The $p_{ks}$-value from the KS test accomplishes both for the linear and nonlinear models. For instance, in Table \ref{tab:lin-and-nonlin-results} under the 2 LST bin case for the 300 MHz BM model input, according to the highest Bayesian evidence, the linear model prefers only the spectral index parameter, while the nonlinear model prefers all three (index, magnitude, and curvature). The linear model is able to produce good fits according to $\chi^2_{red}$ in all parameter cases (all have $\sigma = 1$), while for the corresponding nonlinear models, it is not well-defined whether they correspond to fits at the one or more sigma level. Examining the $p_{ks}$, we find that the two nonlinear cases pass the null hypothesis, but are not preferred over the highest $p_{ks}-$value of the linear model, \textit{even though both nonlinear cases have higher evidences}.

Another interesting example is the 5 LST bin, (300/130) PM model case, which illustrates the possible failure of the evidence to pick a preferred and good-fitting model: here, the nonlinear model with all three parameters (spectral indices, magnitudes, and curvatures) is preferred by the evidence compared to the corresponding linear model, and yet if one calculates the nonlinear model $\chi^2_{red}$ na\"{i}vely using $N_{\theta}$ as the degrees of freedom, one finds $\chi^2_{red} = 7.9$, which is almost certainly not a good fit. The $p_{ks}-$value, however, sorts out the difficulty, as the linear model has $0.93$, which is orders of magnitude higher than the nonlinear value of $8.6 \times 10^{-20}$. The linear model's $\chi^2_{red}$ value further confirms this preference.

Remarkably, the linear model does not produce any fits that fail the null hypothesis until using 10 LST bins, whereas the nonlinear model fails in many instances before that when the model input is incorrect and there are not sufficient nonlinear parameters included (for our method of LST-binning). At 10 LST bins, no nonlinear model produces good fits for any of the inputs described, as shown by the graying-out of all the nonlinear rows. The linear model, on the other hand, exhibits several good-fitting models, all of which require the complexity added by including all three parameters. The model input cases corresponding to the incorrect patch map labelled (300/130) PM appear to be the most difficult cases for the nonlinear model to fit, and for 5 LST bins the nonlinear model is already incapable of producing any good fits. The number of LST bins, or time-dependence, that the nonlinear model can accommodate is largely limited by the size of the beam, and therefore the number of distinguishable patches on the sky (see Appendix \ref{app-skyregions} for a brief discussion on the effect of beam size on the number of regions and LST bins preferred by the evidence).
\begin{figure*}
    \centering
    \includegraphics[width=0.96\textwidth]{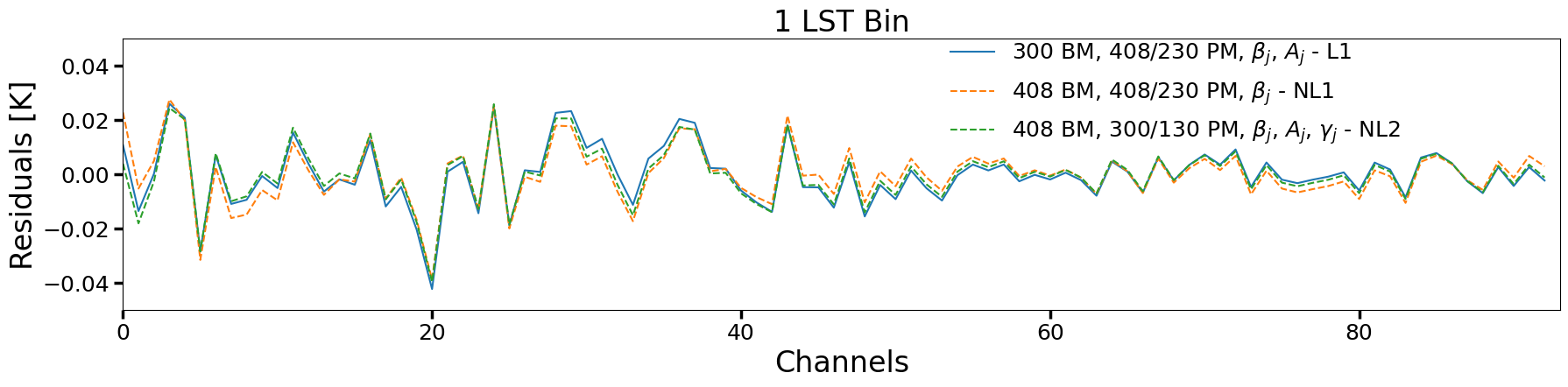}
    \includegraphics[width=0.96\textwidth]{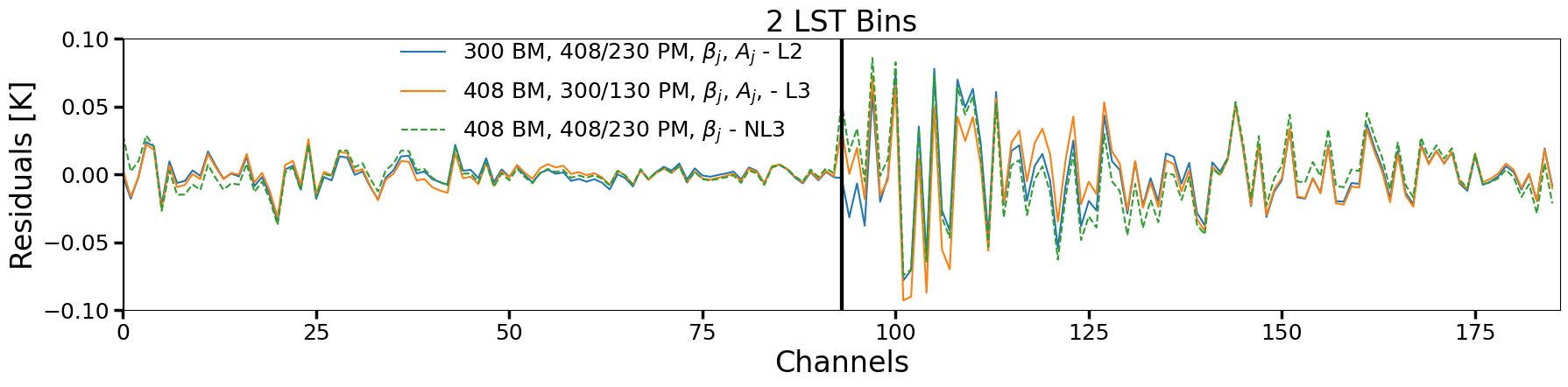}
    \includegraphics[width=0.96\textwidth, height=6cm]{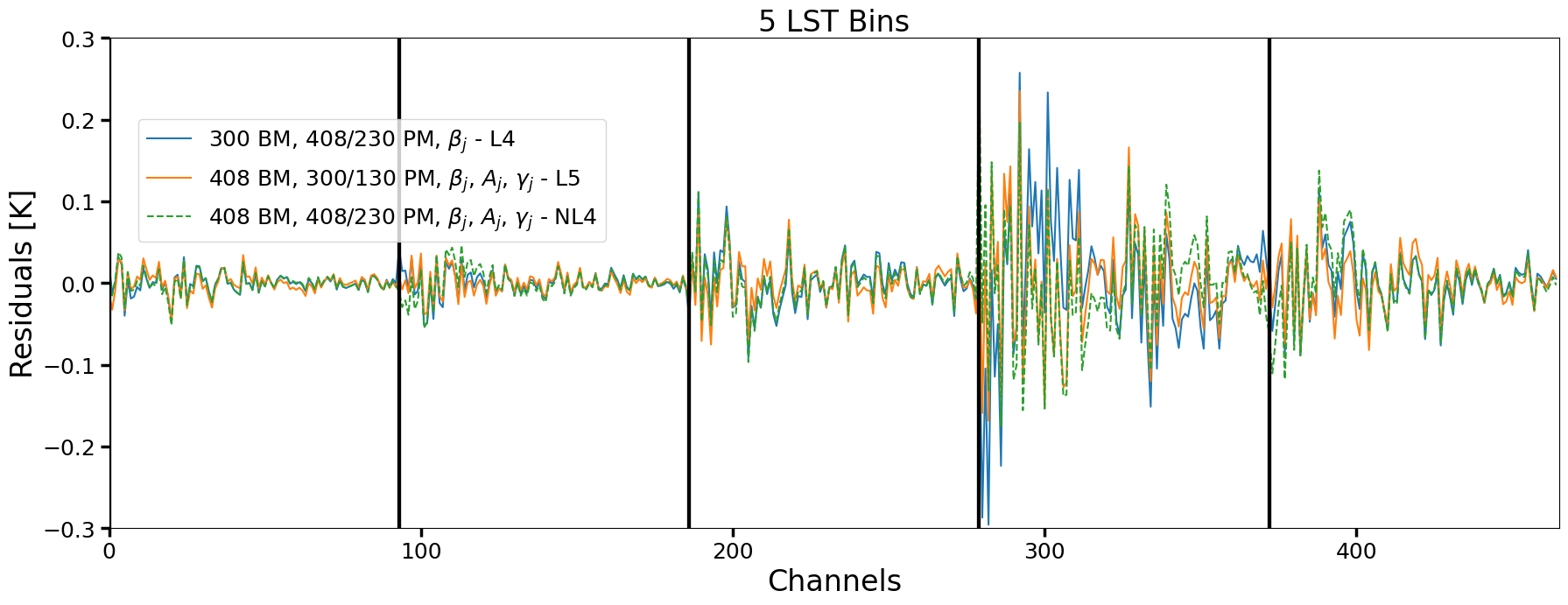}
    \includegraphics[width=0.96\textwidth]{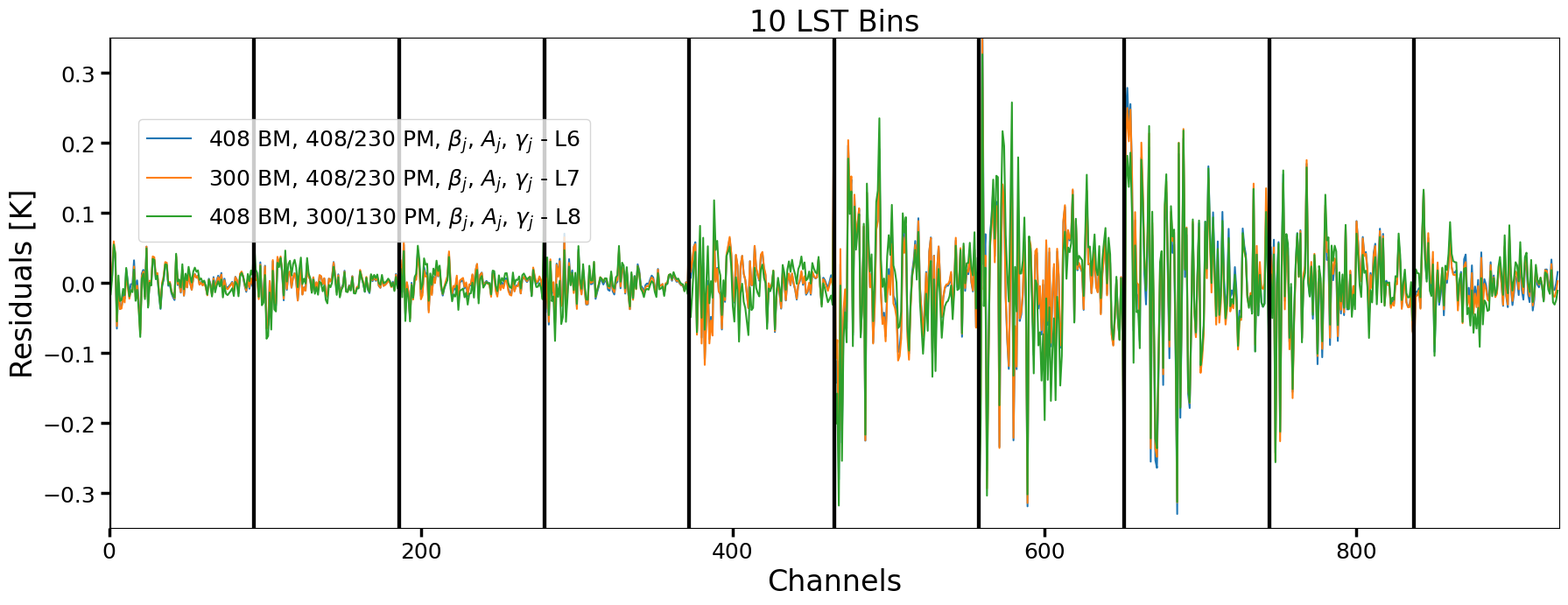}
    \caption{Residuals from the best-fitting (BF) cases from Table \ref{tab:lin-and-nonlin-results} for the linear and nonlinear models. Each panel is denoted by the number of LST bins in the fit. Nonlinear models use dotted lines and have $NL$ in their label, while linear models use solid lines and are denoted by $L$.}
    \label{fig:nl-and-lin-residuals}
\end{figure*}

Figure \ref{fig:nl-and-lin-residuals} shows the residuals produced for each of the best-fit linear and nonlinear models. Best fit models that are linear are shown by solid lines, while those that are nonlinear are shown by dotted lines. For the 1 and 2 LST bin cases, it is clear that the best-fit models remove all foreground power, leaving only the input Gaussian-distributed noise. However, for both the 5 and 10 LST bin cases, especially in the channels where the power is greater (because the galaxy is overhead), even the best-fit models leave small amounts of residual foreground power which manifests as small ripples in the spectrum (see for instance the model L8 represented by the green line in 10 LST bins within the 400 to 900 channel band). These ripples can be described by sinusoidal functions, as we have verified by taking the Fourier transform of each LST bin and finding that the residual power contains features with periodicity of $\sim 0.025$ cycles/MHz, although in this case they are below the noise level. In particular, we found that the models which do not produce good fits (via the KS-test) produced ripples with considerable power, in the case of the nonlinear model for 2, 5, and 10 LST bins. In contrast, for the linear model the ripples are below the noise level when the null hypothesis is satisfied for all LST bins. 

\subsection{Polynomial Models}

Table \ref{tab:poly-fits} gives the statistics for the fits of the polynomial models to the REACH mock. Both the LinLog and the Lin Polynomial fail the KS test for $N_{py} = 5$ parameters, a number of terms commonly employed in the literature for fits to galactic foreground spectra. $N_{py} = 6$ passes the KS-test, however, and while it is still three sigma away from the expected $\chi^2_{red}$, it thus produces residuals which are indistinguishable from noise-normalized foreground-free residuals. Polynomial models with $N_{py} = [6,7,9]$ have comparable evidences, but we find that $N_{py} = 9$ is preferred because of its low sigma value $\sigma = 2.1$ compared to the other two models, and it also exhibits the highest $p_{ks}$-value of the polynomial fits.

We find that at least $N_{py} = 10$ is required to produce fits with $\sigma \approx 1$ and RMS residuals on the order of $10$ mK. Furthermore, we find that these models are \textit{not} preferred, according to the highest $p_{ks}-$value for 1 LST bin cases, when compared to the 408 BM linear and nonlinear models. The LinPhys Polynomial model fails outright to fit the REACH mock ($\chi^2_{red}$ = 9.04), and would require many more terms to fit out any residual foreground power.

\begin{figure*}%
    \centering
    \subfloat[\centering Polynomial Models.]{{\includegraphics[width=9cm]{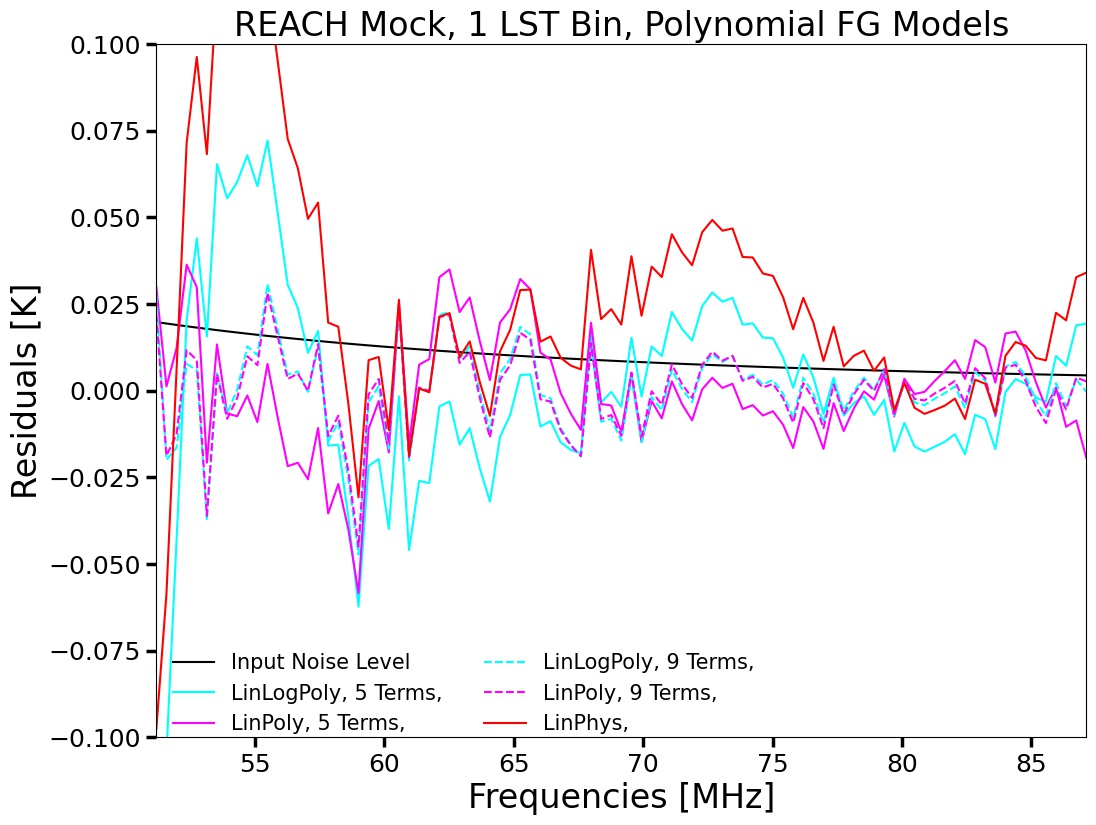}}}%
    \subfloat[\centering Maximally Smooth Polynomial Models.]{{\includegraphics[width=9cm]{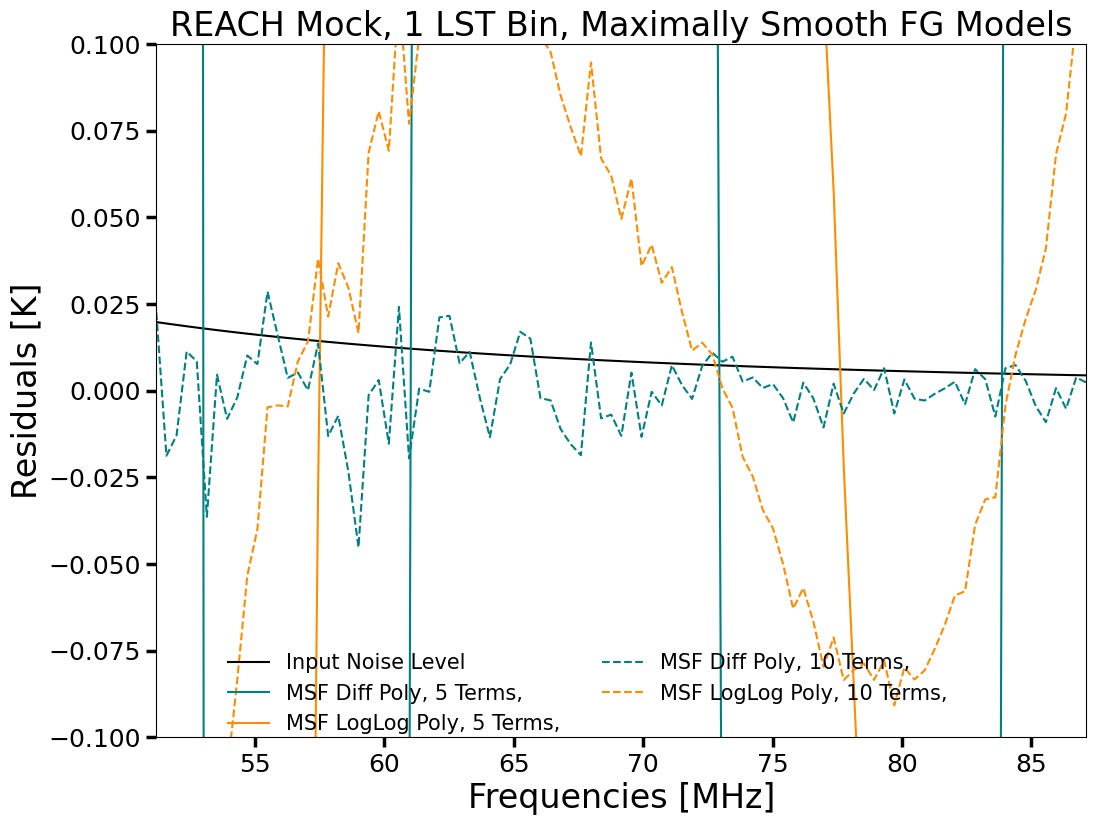}}}%
    \caption{Panel (a) shows several example residuals produced from fitting polynomial models to the REACH mock. Solid (dashed) lines denote 5 (9) term fits. The input noise level is shown as a solid black line, while the red, cyan, and magenta lines denote the linear physical, linear-logarithmic, and linear polynomial models, respectively. Panel (b) shows similar example residuals for the maximally smooth polynomials, with solid (dashed) lines for 5 (10) parameter fits. Blue denotes the difference polynomial and orange the log-log polynomial model.}%
    \label{fig:polynomial-msf-residuals}%
\end{figure*}

Panel (a) of Figure \ref{fig:polynomial-msf-residuals} shows the residuals produced by various polynomial model fits to the REACH mock, including the best-fit cases with 9 terms. It is interesting to note that increasing the number of terms in the fit merely reduces the residual amplitudes, but does not appear to alter their structure significantly. That is, the ripples in the residuals follow a similar sinusoidal pattern with roughly the same period for the LinLog and Lin Polynomial models even when including more terms. 

To verify the robustness of the above results, we conducted additional fits to mock foregrounds where we allowed the BM $T_0$ used for extrapolation and the spectral index map $\beta_{R0}$ generating the sky temperatures to be different from the REACH mock. We both fixed $T_0 = 408$ MHz, and changed the spectral index map to $\beta_{R0} = (408/300), (300/100), (408/100)$, and also fixed the spectral index map to $\beta_{R0} = (408/230)$ and changed $T_0 = 300, 200, 45$ MHz, giving six new intrinsic foregrounds that we then convolved according to the same beam and observation strategy. We found that all mock foreground cases required $N_{py} > 5$ in order to achieve good fits. Therefore, we conclude that the polynomial models, for realistic experiments, require $N_{py}>5$ for good fits, and typically the best fits have $N_{py} \sim 9$ terms, although $N_{py} = 6$ is not disallowed according to our tests.

\subsection{Maximally Smooth Polynomials}
Lastly, Table \ref{tab:msf-fits} shows the statistics and model fits produced when fitting the REACH mock with the two maximally smooth polynomials. We find that the LogLog model is unable to fit out enough of the foreground power to pass the null hypothesis, even with $N_{MSF} = 15$ terms. On the other hand, the Diff model requires at least $N_{MSF} = 10$ terms to achieve a good fit. A greater number of terms, however, seems to diminish the fit, as seen by the $N_{MSF} = 15$ terms case, and as we verified by calculating the fits for $N_{MSF} = 1 - 20$. We conclude that the maximally smooth polynomials in general require higher numbers of terms to fit the REACH mock than the polynomial models and moreover are not preferred over the latter (as per the higher of the various $p_{ks}-$values).

Panel (b) of Figure \ref{fig:polynomial-msf-residuals} shows the residuals produced by several maximally smooth polynomial cases, which are clearly much larger than their polynomial counterparts for a given number of terms -- compare them to those in panel (a) in the same scale. The poor fits produce similar sinusoidal ripples which are indicative of unmodelled foreground power, though they have a different residual periodicity than the polynomial models.

Lastly, we note that the above analyses for the polynomial and maximally smooth polynomials neglect the potential added fitting benefit of multiplying the spectrum by a beam-chromaticity correction, such as in \citep{bowman_absorption_2018,sims_bayesian_2023}. In particular, the latter work shows how such a factor removes chromaticity coupling perfectly in a spectrum with a spatially isotropic intrinsic foreground, or can dampen residual foreground power when the spatially anisotropic intrinsic foreground can be described by perturbations which are constructed from LinLog polynomials. We leave an investigation of how such a correction factor might affect our own beam-weighted foreground fits to future work.

\section{Discussion}
\label{sec-discussion}

\subsection{Nonlinear Model: Reconstructed Intrinsic Foreground}
\begin{figure*}%
    \centering
    \subfloat[\centering RMS Error of the Best-fit Nonlinear Foreground Models Compared to the Input Intrinsic Foreground.]{{\includegraphics[width=9cm]{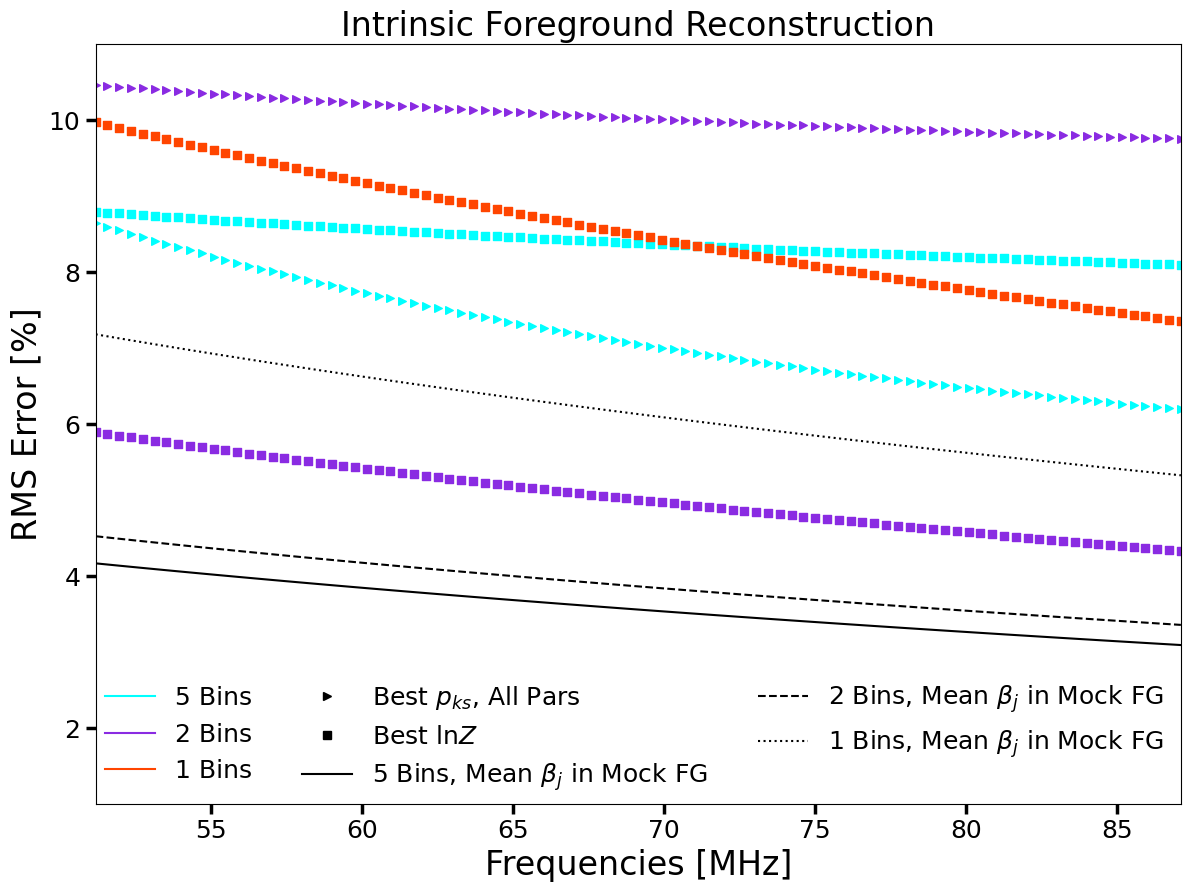}}}%
    \subfloat[\centering Reconstructed foreground as a function of LST.]{{\includegraphics[width=9cm]{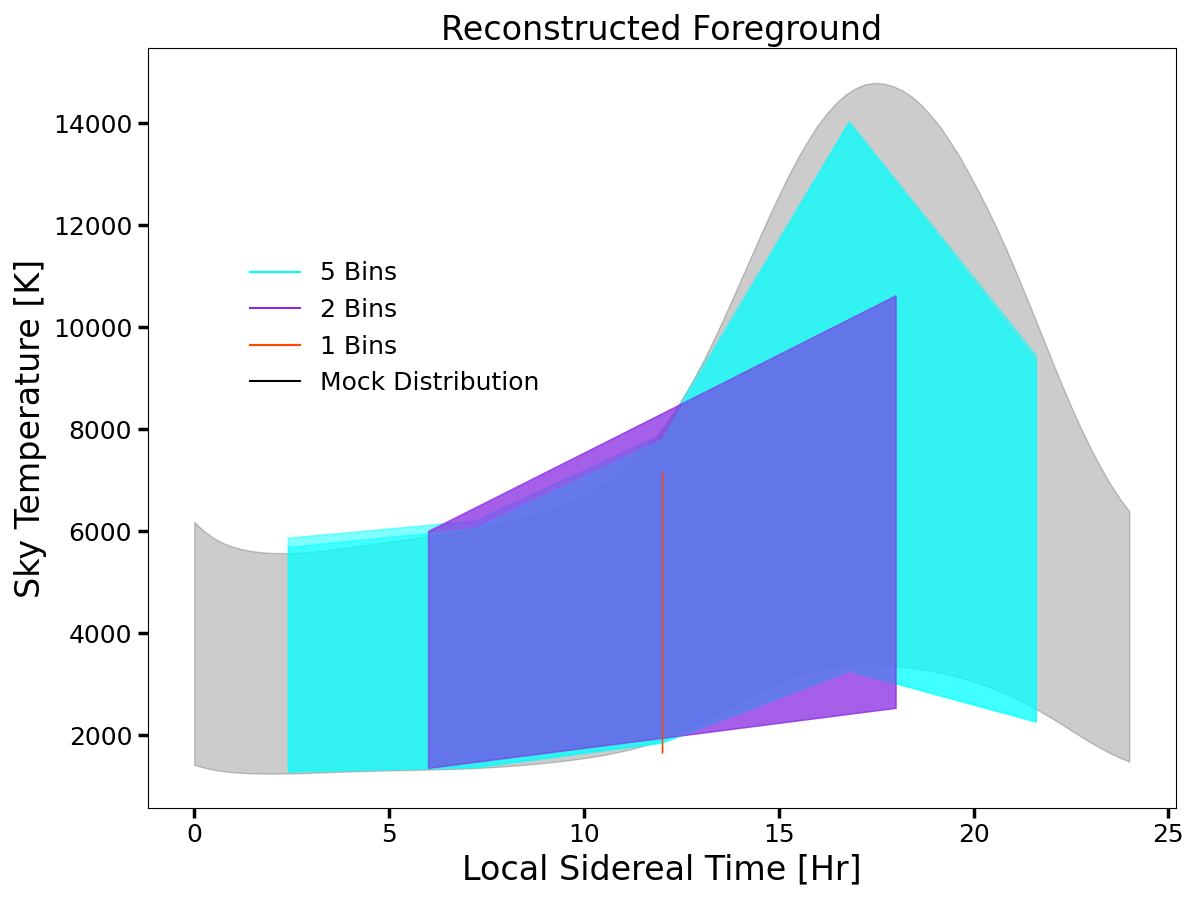}}}%
    \caption{A comparison of the reconstructed intrinsic foreground as determined by the best-fit nonlinear parameters from fits to the convolved spectra. The left panel shows the RMS error in the reconstruction (as computed with respect to the input intrinsic foreground) as a function of frequency, where the different colors (red, purple, and cyan) label the different LST bin cases (1, 2, and 5 respectively). The triangle symbols label models which were selected as best-fit via the KS-test, whereas the squares denote models selected using the Bayesian evidence. The black lines correspond to the ideal case, without the beam or averaging over the sky, by using the mean spectral index in the input intrinsic foreground for each spectral region, with the number of regions $N_r=$ 4, 8, and 9, for 1, 2, and 5 LST bins, respectively. The right panel shows the reconstructions of the foreground as a function of LST for the models selected via the KS-test. The gray band shows the input intrinsic foreground. The top (bottom) of the band corresponds to the lowest (highest) frequencies.}%
    \label{fig:intrinsic-fg-recon}%
\end{figure*}

\begin{figure*}
    \centering
    \includegraphics[width=0.96\textwidth]{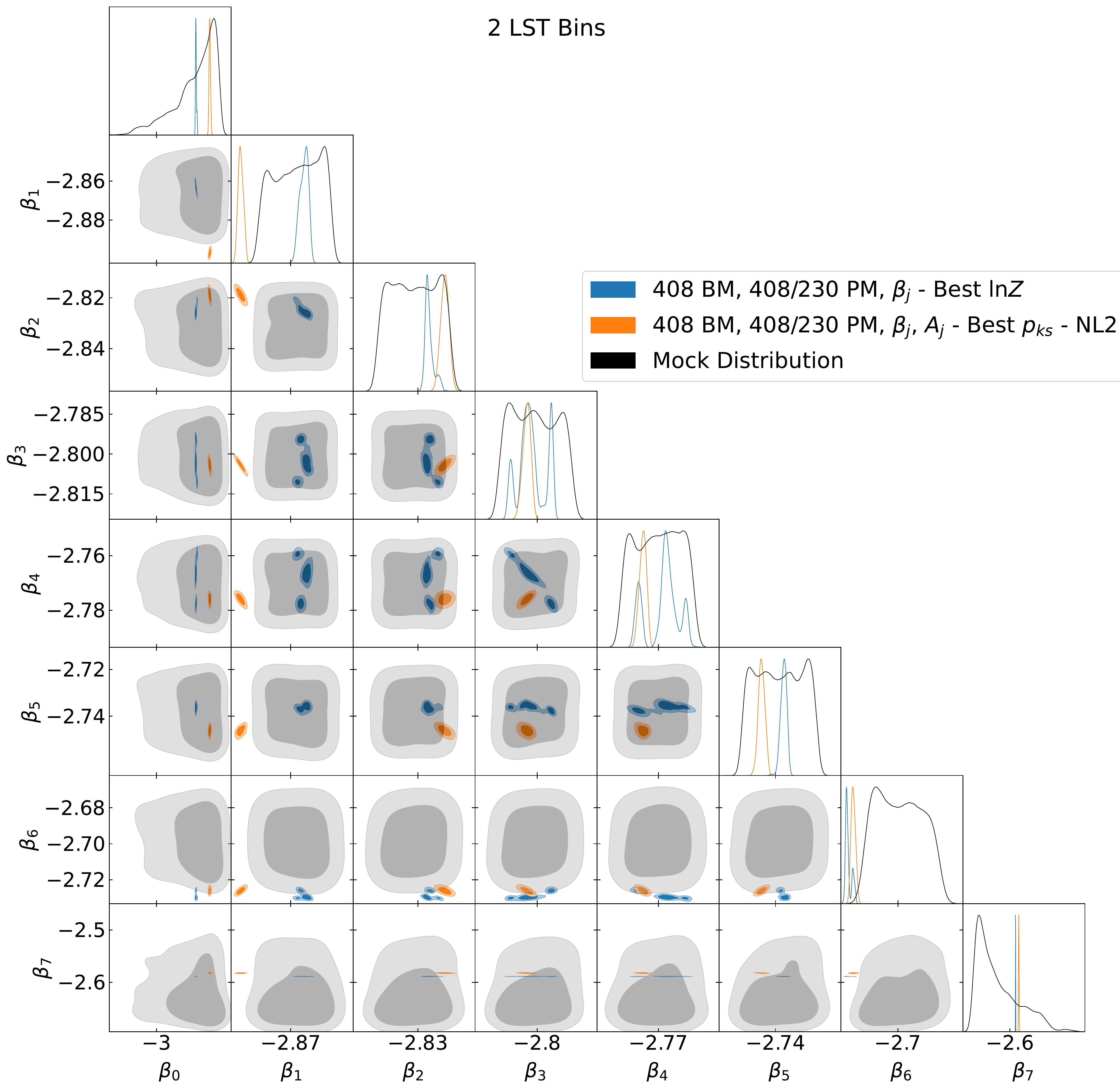}
    \caption{Example triangle plot for the 2 LST bin case, showing the best-fit spectral indices for each region, $N_r = 1 - 8$. The blue contours correspond to the best-fit model selected by the evidence, while the orange contours correspond to the model selected by the KS-test. The black contours show the input mock distribution for the intrinsic foreground, as generated from the GSM. Contours which are closer to the center of the mock distribution represent better fits to the intrinsic foreground, but not necessarily to the spectra which result from convolving the foreground with a beam.}
    \label{fig:triagnle-plot-betas}
\end{figure*}

The nonlinear model allows us to reconstruct an intrinsic foreground from the convolved mock spectra, as seen through the beam and sky-averaging. This is important because, due to the latter two effects, there is no a priori reason that the nonlinear parameters, such as the spectral index, will correspond exactly to those in the input intrinsic foreground. Moreover, there is a fundamental limitation on how well the nonlinear model can fit the intrinsic foreground, even neglecting the beam and sky-averaging, due to the fact that we are approximating a sky with tens of thousands of pixels by effectively ten pixels (i.e. regions) or less.

In Figure \ref{fig:intrinsic-fg-recon} we show the intrinsic foreground reconstructions for the best-fit nonlinear models with 1, 2, and 5 LST bins. For 10 LST bins there were no fits that passed the null hypothesis test. In the left panel of the figure we show the RMS of the difference---or error---between the REACH mock intrinsic foreground and the reconstructed foreground as a function of frequency. That is, at each frequency we reconstruct an intrinsic foreground using the best-fit nonlinear parameters as found from the fit to the corresponding convolved mock spectra, and the following equation for the intrinsic, reconstructed foreground, $T_{\text{IRFG}}(\Omega, \nu)$:
\begin{equation}
    T_{\text{IRFG}}(\Omega, \nu) = \sum^{N_r}_{j=1} (T_0 - T_{\text{CMB}}) A_j \left( \frac{\nu}{\nu_0} \right)^{\beta_j} \left( \frac{\nu}{\nu_0} \right)^{\gamma_j \ln{\nu/\nu_0}},
\end{equation}
where the number of regions $N_r$ and nonlinear parameters are determined by the particular fit (See \citep{chluba_rethinking_2017} however for a much more general formula using spherical decompositions and various foreground components). If a model fit does not have magnitude or spectral curvature parameters, we set them to unity or zero, respectively. 

In the left panel of Figure \ref{fig:intrinsic-fg-recon}, the different colors correspond to the different LST bins, while the shapes correspond to whether we use the $p_{ks}-$value to determine the best-fit model (triangles), or the evidence (squares). The black dotted, dashed, and solid lines show the RMS error when we use the mean value of the spectral index $\beta_R$ in each region of the REACH mock to reconstruct the foreground, and thus they are not affected by the beam or sky-averaging. Given the latter ideal circumstances, the black lines represent in essence the fundamental limit of the model to fit the data. When approximating a sky with many pixels by a few regions (4, 8, and 9 for 1, 2, and 5 LST bins, respectively), one cannot do better than an RMS error of roughly four percent, as shown by the solid black line for 5 LST bins. 

Interestingly, many of the best fit models (colored lines) produce RMS errors with respect to the intrinsic foreground that are relatively close to the ideal case (black lines), and overall roughly between 5 to 10 percent. It is important to remember that for these cases we did not fit the intrinsic foreground, but the convolution of it with the beam, so once again, we do not expect to obtain parameter values that ideally fit the intrinsic foreground. The evidence and KS-test both choose the same model for the 1 LST bin case, namely that of only spectral indices, which is why their symbols overlap and there appears to be only one red line in the left panel of the figure. The 5 LST bin cases (in cyan) both give around 8 percent error, while there is a slightly larger separation between the two purple lines for 2 LST bins with different model fit statistics (see also Figure~\ref{fig:triagnle-plot-betas} below). This is because the best-fit model as selected by the KS-test for the 2 LST bins incorporates magnitude parameters (triangle symbols), while the other fit has only spectral indices (circle symbols).

The right panel of Figure \ref{fig:intrinsic-fg-recon} shows the reconstructed foreground (colored regions) as a function of LST. Each shaded region encapsulates all frequencies, with the top of the band corresponding to the lowest frequencies, and the bottom of the band, the highest frequencies. The gray region shows the input mock spectra as a function of LST. Clearly, using more time-dependence information improves how well we can approximate the whole sky, although there is a limit to how many LST bins can be used before the nonlinear model breaks down, as noted above.

Lastly, in Figure \ref{fig:triagnle-plot-betas} we show a triangle plot of the spectral index parameters for the two best-fit models (as determined by the evidence and KS-test, respectively) for 2 LST bins. As before, the input mock spectral index distributions are outlined in black in the one-dimensional panels, and appear as gray contours in the 2-dimensional panels. These distributions represent again the ideal case of intrinsic foreground reconstruction, so the closer the blue (best evidence case) and orange (best KS-test case) distributions are to the centers of the black/gray distributions, the better the reconstruction. Note that the model corresponding to the orange contours has magnitude parameters, while the blue does not. Note also that even for the worst case for the orange contours ($\beta_1$), the reconstructed values are only off from the mean value of roughly $-2.87$ by $\sim0.02$, well within real observational errors \citep{guzman_all-sky_2011,mozdzen_spectral_2019}. Therefore, our reconstruction of the intrinsic foreground using best-fit nonlinear parameters from a fit to the convolved mock spectra is quite encouraging, despite the corruption induced by the chromatic beam.

\subsection{P-values in the KS-test}
\label{sec:p-values}

Astute readers may have noticed that four of the fits for 5 or 10 LST bins in the linear model column produced $p_{ks} > 0.05$ and thus passed the null hypothesis test, but according to the analytical calculation of $\chi^2_{red}$, still had $\sigma > 2$. In particular, the last row has a $p_{ks} = 0.12$ but is $\sigma = 7$ away from the $E[\chi^2_{red}]$. When the data vector is large, as in those cases, a higher threshold for the null hypothesis test would alleviate such apparent tension.~By choosing a threshold p-value of $0.5$, for instance, a ten-fold increase, some of the models that are currently not, would be colored gray. So the p-value threshold may be made more stringent, in accordance with the nature of the test and data vector. Note also that a higher number of LST bin data vectors for the same amount of integration time implies a greater noise level in each bin. 

\section{Conclusions}
\label{sec-conclusions}
There are two primary sources of bias and error in 21-cm signal extraction: the first comes from overlap between signal and foreground models, while the second comes from inadequate foreground models alone which cannot fit mock spectra to the noise level of the data required for signal extraction. In this work, for a given mock foreground spectrum, we tested seven galactic foreground models commonly employed for 21-cm cosmology, including a physically-motivated nonlinear model, a linear model characterized by the eigenmodes of the latter, three polynomials, and two maximally-smooth polynomials in order to determine what residual level, and hence potential bias, they produce, given different model input assumptions and their number of foreground parameters to generate the mock data.

We first simulated mock foreground spectra that incorporated realistic spatial and spectral complexity, beam models, horizon profiles, discrete time-sampling, and low noise levels for different numbers of LST bins, for an EDGES-like experiment. Next, we fitted each of the seven models to the mock foreground spectra for various numbers of parameters and input model assumptions, including the dependence of the forward-models upon their input spatial temperature and spectral index maps. Lastly, we calculated the Bayesian evidence and two goodness-of-fit statistics for each fit: $\chi^2_{red}$ and the p-value, $p_{ks}$, from a KS-test comparing the noise-normalized residuals of the fit to a unit normal distribution. We recommend using the latter statistic in particular as it denotes goodness-of-fit and model preference between two models when the degrees of freedom are unknown or indeterminable (as in the case of nonlinear models), as opposed to the traditionally used $\chi^2_{red}$. We found the following results:

\begin{itemize}
    \item For 1 LST bin, to which all the models were applied, the nonlinear model was preferred according to the $p_{ks}$ statistic ($p_{ks} = 0.99$). This model is able to fit the convolved foreground spectra even when its model spatial inputs, such as the temperature map used for extrapolation or the spectral index map distribution, are different from the mock spectrum, needing 4 parameters in the ideal case and 12 in the worst. On the other hand, the linear model fits the mock spectra well in all cases with about 6-7 parameters ($p_{ks} = 0.94,0.97$ respectively), making it only slightly less preferred than the nonlinear model.
    \item For 1 LST bin, the polynomial and maximally smooth polynomial models do not provide good fits when using only 5 parameters $p_{ks} < 10^{-6}$, as commonly used in the literature for these models. However, we do find that 6 parameters produces a fit which passes the null hypothesis (residuals are indistinguishable from foreground-free residuals) for the linear, logarithmic polynomial model $p_{ks} = 0.4$, and that it is only three sigma away from the expected reduced chi-squared with a Bayesian evidence comparable to the 7 and 9 term polynomial fits $\Delta \ln{Z} \leq 1$. However, both the linearized physical polynomial and the log-log maximally-smooth polynomial models could not fit the mock spectra with any number of parameters.
    \item When jointly fitting multiple LST bins, we found that the linear model significantly out-performed the nonlinear model (according to the $p_{ks}$ value and our choice of LST-binning). The nonlinear model begins to leave foreground power in the residuals starting at 2 LST bins, and fails goodness-of-fit tests ($p_{ks}<0.05$) utterly in every case by 10 LST bins. In contrast, the linear model only begins to leave residual foreground power at 10 LST bins, and even then still achieves good fits when the model is made sufficiently complex, even when the input spatial temperature map or spectral index map are different from those used to generate the mock. The linear model is also orders of magnitude faster than its nonlinear counterpart.
    \item The $p_{ks}$-value is a more robust and general statistic to use for model selection and goodness-of-fit than the Bayesian evidence and reduced chi-squared $\chi^2_{red}$ when comparing different foreground models. For nonlinear models in particular, we found several instances where the nonlinear model had high evidence $(-234)$ compared to the linear model $(-1130)$, but no unambiguous $\chi^2_{red}$ goodness-of-fit, as one cannot accurately calculate the degrees of freedom for a nonlinear model. Examining the $p_{ks}$-values, however, clears up this ambiguity, giving $0.5$ for the nonlinear model and $0.78$ for the linear model, where the latter is thus clearly preferred.
\end{itemize}

It is also worth recalling here that for joint global signal plus foreground fits, the linear foreground model is also a desirable choice as its parameters can be efficiently and analytically marginalized over \citep{rapetti_global_2020, tauscher_global_2021}. In addition, the overlap between the linear foreground and global signal models is significantly decreased by employing more LST bins that are modeled simultaneously \citep{tauscher_global_2020}, making the linear model desirable for joint analyses. This method of decreasing the overlap between foreground and global signal models increases the number of linear foreground parameters, which is largely how good fits to the mock foreground spectra were obtained in this paper. In contrast, for the other foreground models, the nonlinear and the two types of polynomials, increasing the number of foreground parameters to achieve good fits inevitably leads to more overlap with global signal models.

\section{acknowledgements}
We would like to thank Steven Murray for useful discussions and advice on the noise estimation and modelling. We would also like to thank Keith Tauscher for his help with the analytical evidence calculations, and Jordan Mirocha, Dominic Anstey, and Peter Sims for helpful discussions and comments. This work was directly supported by the NASA Solar System Exploration Research Virtual Institute cooperative agreement 80ARC017M0006. This work was also partially supported by
the Universities Space Research Association via D.R. using
internal funds for research development. We also acknowledge support by NASA grant 80NSSC23K0013. This work utilized the Blanca condo computing resource at the University of Colorado Boulder. Blanca is jointly funded by computing users and the University of Colorado Boulder.

\appendix 

\section{Analytical Calculation of the Evidence}
\label{app-evidence}

As the likelihood does not change for a given data vector, we can neglect its normalizing factor, as evidence differences (or ratios) cause them to cancel. We must keep the normalizing factor for the prior distribution, however, leading to
\begin{equation}
\begin{split}
    \mathcal{L}(T|x) \propto \exp{-\frac{1}{2} (Fx - T)^T C^{-1}(Fx-T)} \\
    \pi(x) = \frac{1}{\sqrt{(2\pi)^{N_{\Theta}}\abs{\Lambda}}} \exp{-\frac{1}{2} (x - \nu)^T \Lambda^{-1} (x-\nu)}
\end{split}
\end{equation}
where $F$ labels the basis matrix with vectors as its columns, $x$ are the parameters, $T$ is the data, $C$ is the covariance matrix for the data, $\nu$ is the prior parameter mean, and $\Lambda$ is the prior parameter covariance.

Then the evidence integral of Equation \ref{eqn-evidence} becomes
\begin{equation}
\begin{split}
    Z = \frac{1}{\sqrt{(2\pi)^{N_{\Theta}}\abs{\Lambda}}} \int \exp{-\frac{1}{2} (Fx - T)^T C^{-1}(Fx-T)} \exp{-\frac{1}{2} (x - \nu)^T \Lambda^{-1} (x-\nu)} d^{N_{\Theta}}x \\
    = \frac{1}{\sqrt{(2\pi)^{N_{\Theta}}\abs{\Lambda}}} \int \exp{-\frac{1}{2} [ x^T(F^TC^{-1}F + \Lambda^{-1})x -2(F^TC^{-1}T + \Lambda^{-1} \nu)^Tx + (T^TC^{-1}T + \nu\Lambda^{-1}\nu)]} d^{N_{\Theta}}x
\end{split}  
\end{equation}
where we have collected the terms in powers of $x$. Those with no dependence on $x$ come outside the integral, and we recognize the term quadratic in $x$ as nothing more than $S^{-1}_{\text{MAP}}$, the best-fit parameter covariance. We also define $\mu\ \equiv F^TC^{-1}T + \Lambda^{-1}\nu$ to substitute for the term linear in $x$, giving:
\begin{equation}
    Z = \frac{e^{-\frac{1}{2}(T^TC^{-1}T + \nu\Lambda^{-1}\nu)}}{\sqrt{(2\pi)^{N_{\Theta}}\abs{\Lambda}}} \int \exp{-\frac{1}{2} [x^T S^{-1}_{\text{MAP}} x -2\mu^T x]} d^{N_{\Theta}}x.
\end{equation}
We introduce the new variable $y = S^{-1/2}_{\text{MAP}}x$ and substitute to get
\begin{equation}
\begin{split}
    Z = \frac{e^{-\frac{1}{2}(T^TC^{-1}T + \nu\Lambda^{-1}\nu)}}{\sqrt{(2\pi)^{N_{\Theta}}}} \sqrt{\frac{\abs{S_{\text{MAP}}}}{\abs{\Lambda}}} \int \exp{-\frac{1}{2} [y^Ty -2\mu^T S^{1/2}_{\text{MAP}}y ]} d^{N_{\Theta}}y \\
    = \frac{e^{-\frac{1}{2}(T^TC^{-1}T + \nu\Lambda^{-1}\nu)}}{\sqrt{(2\pi)^{N_{\Theta}}}} \sqrt{\frac{\abs{S_{\text{MAP}}}}{\abs{\Lambda}}} \int \exp{-\frac{1}{2} [(y-S^{1/2}_{\text{MAP}} \mu)^T(y-S^{1/2}_{\text{MAP}} \mu) - \mu^T S_{\text{MAP}} \mu]} d^{N_{\Theta}}y
\end{split}
\end{equation}
where we have completed the square. Defining $Q \equiv y - S^{1/2} \mu$ and bringing terms outside the integral which are not dependent on $y$, we find
\begin{equation}
    Z = \frac{e^{-\frac{1}{2}(T^TC^{-1}T + \nu\Lambda^{-1}\nu - \mu^T S_{\text{MAP}} \mu)}}{\sqrt{(2\pi)^{N_{\Theta}}}} \sqrt{\frac{\abs{S_{\text{MAP}}}}{\abs{\Lambda}}} \int \exp{-\frac{1}{2} Q^TQ} d^{N_{\Theta}}Q
\end{equation}
where the integrand is equal to $(2\pi)^{N_{\Theta}/2}$, leaving us with
\begin{equation}
    Z = \sqrt{\frac{\abs{S_{\text{MAP}}}}{\abs{\Lambda}}} \exp{\frac{\mu^T S_{\text{MAP}} \mu - T^TC^{-1}T - \nu\Lambda^{-1}\nu}{2}},
\end{equation}
as presented in Equation \ref{eqn=anal-lin-ev}. 

\section{Time dependence, sky regions, and beam size}
\label{app-skyregions}

\begin{figure*}
    \centering
    \includegraphics[width=0.96\textwidth]{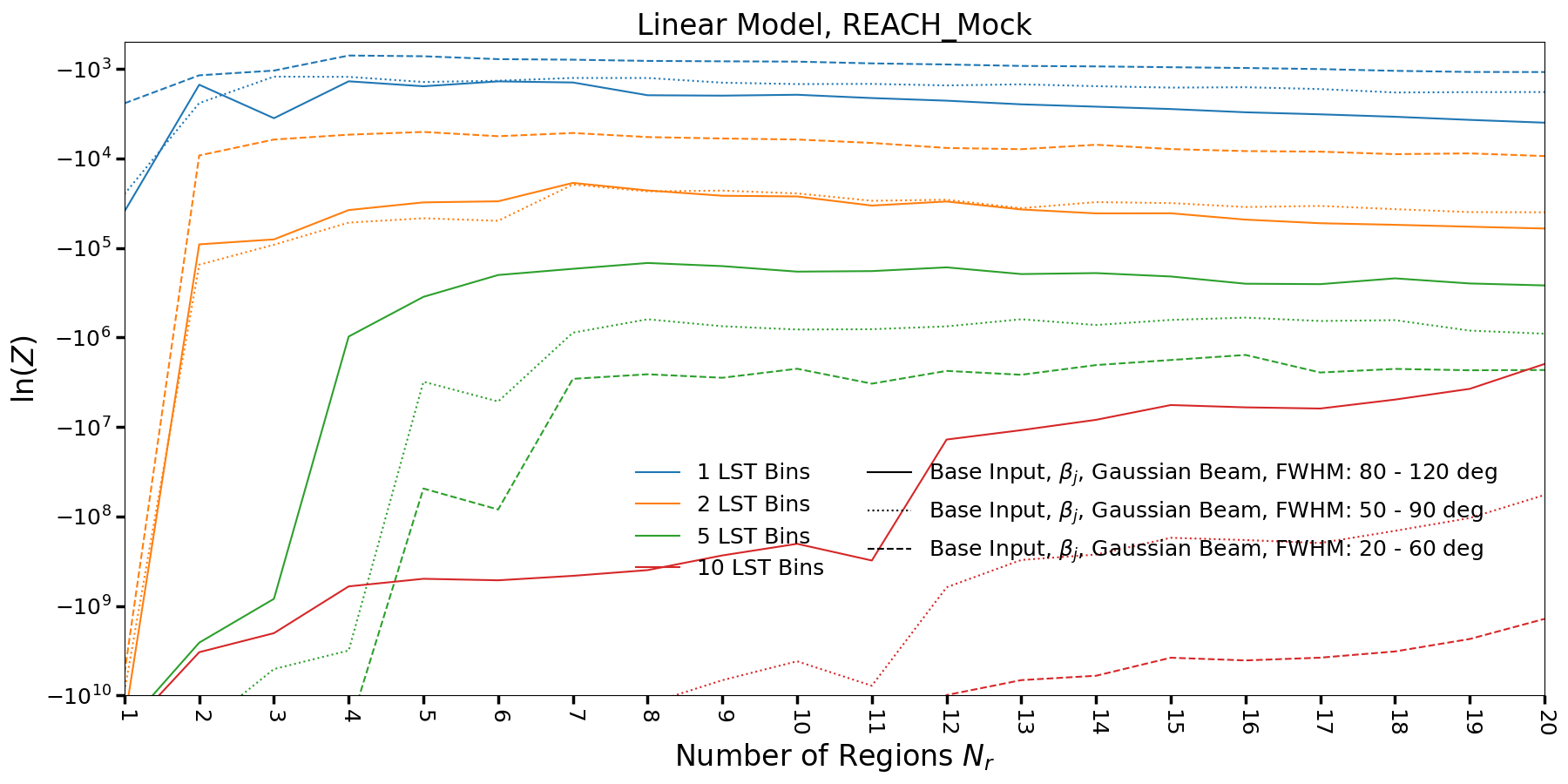}
    \caption{Bayesian evidence as a function of the number of regions of the sky $N_r$ and the number of LST bins (compare to the top panel of Figure \ref{fig:lin-model-evidence-chi}). For each line-style, the legend shows the relative size of the FWHM of the analytical Gaussian beam as a quadratic function of the band. The first number refers to the degree FWHM of the beam in the low band, while the second number the shows the high band. The dotted and dashed lines therefore have smaller beams across the band than the solid line. For 1 and 2 LST bins (blue and orange contours), the best-fit number of regions saturates to the same value of $N_r$ quickly, regardless of the size of the beam. However, when the number of LST bins increases to 5 or 10, the best-fit (according to the highest evidence) number of regions $N_r$ increases dramatically for beams which have smaller FWMHs across the band (dotted and dashed lines).}
    \label{fig:beamsize_vsregions}
\end{figure*}

Figure \ref{fig:beamsize_vsregions} shows the Bayesian evidence as a function of the number of regions that the sky is split into for fits with a beam which is spatially Gaussian and with a Full-Width Half-Maximum (FWHM) that varies as a quadratic polynomial across the band. Different colors show different numbers of LST bins fit, and hence we expect greater time-resolution to require higher spatial resolution---or more regions---which is indeed what the figure shows. Moreover, a larger beam size averages more of the sky together, making finer resolution sky regions indistinguishable in the convolved spectra. The solid line shows the case for a beam which has a FWHM of 80 degrees in the low band, and varies up to 120 degrees in the high band. For comparison, the dotted and dashed lines show beams which have FWHMs which are 30 and 60 degrees smaller across the band compared to the solid line. Clearly decreasing the size of the beam increases the number of distinguishable sky regions, and hence also increases the time-resolution of the forward-model.

\bibliographystyle{aasjournal}
\bibliography{zotero_library_ref}

\end{document}